\documentclass[11pt]{article}

\usepackage[a4paper,top=3 cm,bottom=3 cm,inner=2.5 cm,outer=2.5 cm]{geometry} 

\usepackage{amsmath,amssymb,amsfonts,mathrsfs} 

\usepackage{graphics,graphicx}
\usepackage{amscd}

\def\bb{\begin{eqnarray}}
\def\ee{\end{eqnarray}}

\newcommand{\Att}{\mbox{\large \tt{A}}}
\newcommand{\Btt}{\mbox{\large \tt{B}}}
\newcommand{\Ctt}{\mbox{\large \tt{C}}}
\DeclareMathAlphabet\mathbfcal{OMS}{cmsy}{b}{n}

\newcommand{\Proof}{\begin{proof}}
\newcommand{\QED}{\end{proof} \noindent}

\newcommand{\Mb}{{\boldsymbol{M}\hspace*{-13.15pt}\boldsymbol{M}}}

\begin{document}
${}$
\begin{center}

{\Large Lecture Notes on Operator Algebras \\[6pt]
and Quantum Field Theory}
\\[12pt]
EMS--IAMP Spring School\\[4pt] ``Symmetries and Measurement in Quantum Field Theory''\footnote{The Spring School took place at the University of York, UK, April 7-11, 2025, organized by C.J.\ Fewster, D.W.\ Janssen and K.\ Rejzner, and funded by EPSRC Grant EP/Y000099/1 to the University of York, the European Mathematical Society, the International Association of Mathematical Physics, and COST Action (European Cooperation in Science and
Technology) CA23115: Relativistic Quantum
Information. The material presented in these notes is an expanded version of the material presented during the lectures by the author. The author would like to thank the participants of the Spring School for their feedback and for discussions.} 
\\[16pt]
{\large  {\bf Rainer Verch}}
\\[12pt]
Institute for Theoretical Physics, University of Leipzig, 04009 Leipzig, Germany
\end{center}
${}$ 
\\[6pt]
{\bf Abstract} This is a set of lecture notes of four lectures: 1.\ Operator Algebras and Quantum Field Theory, 2.\ Tomita-Takesaki Modular Theory of von Neumann Algebras and the Bisognano-Wichmann/Borchers Theorem, 3.\ Local Covariant Quantum Field Theory, 4.\ Temperature and Entropy-Area Relation of Quantum Fields near Horizons of Dynamical Black Holes. The basic aim is to provide an introduction into some of the contemporary concepts and methods of quantum field theory in the operator-algebraic framework (lectures 1 to 3), and to illustrate (in lecture 4) how they may be applied
in a theme of both perpetual and current interest --- the ``thermodynamics'' (and also, the ``information content'') of quantum matter in the vicinity of black holes. While the topic of ``measurement'' in quantum field theory is not covered in these lectures (this topic has been
presented in the lectures of Chris Fewster [arXiv:2504.17437]), the topic of ``symmetry'' in quantum field theory does make an appearance: On one hand, in 
the form of the geometric action of Tomita-Takesai modular objects associated with certain operator algebras and states as stated in the 
theorems by Bisognano-Wichmann, and by Borchers, and on the other hand, in the form of local general covariance.
\\[6pt]
References are provided at the end of each section. They are organized as follows: There is a list of references directly related to the material of the notes. There is also a list of suggested references for further reading. In both cases, monographs and textbooks are listed first, in alphabetical order by author name. Then articles are listed, again ordered by author name. The references are labelled by author name(s) and publication year.
\newpage \noindent

\section{Operator Algebras and Quantum Field Theory}
\subsection{Operator Algebra Basics}
\setcounter{equation}{0}
Many --- or most --- mathematical formulations of theories of physics use algebraic structures (for several reasons).
\\[4pt]
We start by introducing the structure of an algebra at a very general level, and then successively add more structure.
\begin{itemize}
 \item An \textbf{algebra} $\mathbfcal{A}$ is a vector space (over $\mathbb{C}$) with a \textbf{product}: For any $\Att,\Btt \in \mathbfcal{A}$, there is also a unique product $\Att \Btt \in \mathbfcal{A}$, subject to the rules:
 \begin{align*}
  (\Att \Btt)\Ctt = \Att (\Btt \Ctt)\,, \  \  & \Att(b\Btt + c\Ctt) = b\Att\Btt + c\Att \Ctt\,, \\ 
                      & (a\Att + b\Btt)\Ctt= a\Att\Ctt + b\Btt\Ctt \quad (a,b,c \in \mathbb{C}\,, \ \Att,\Btt,\Ctt \in \mathbfcal{A})
\end{align*}
 \item An algebra $\mathbfcal{A}$ is \textbf{abelian} or \textbf{commutative} if 
 $$ \Att\Btt = \Btt\Att \ \ \ \quad \text{for all} \ \ \Att,\Btt \in \mathbfcal{A} $$
 \item An algebra $\mathbfcal{A}$ is called \textbf{unital} if there is an algebraic \textbf{unit element} $\mathbf{1}$ (unique) satisfying
 $$ \mathbf{1}\Att = \Att \mathbf{1} \ \ \ \quad \text{for all} \ \ \Att,\Btt \in \mathbfcal{A} $$
\end{itemize}

\begin{itemize}
 \item An algebra $\mathbfcal{A}$ is a $*$-algebra if there is a $*$-operation 
 $\Att \mapsto \Att^*$ in $\mathbfcal{A}$ which has the properties 
 $$ \left(\Att^*\right)^* = \Att\,, \ \ (\Att\Btt)^* = \Btt^*\Att^*\,, \ \ (a\Att + b\Btt)^* = \bar{a}\Att^* + \bar{b}\Btt^*\,, \ \ \mathbf{1}^* = \mathbf{1} $$ 
 for all $a,b \in \mathbb{C}$ and $\Att,\Btt \in \mathbfcal{A}$.
\end{itemize}
Unital $*$-algebras are prominently used in the mathematical formulation of physical theories. 
In this context, it is useful to have concepts of continuity/topology for such algebras.
A topology for an algebra should be such that the vector space operations, the algebra product and the $*$-operation are continuous, and that the algebra is complete in the topology.
\\[4pt]
Since an algebra $\mathbfcal{A}$ is a vector space, it can usually be endowed with a norm $||\,.\,||$. An \textbf{algebra norm} has the property
$$ ||\Att\Btt|| \le ||\Att||\,||\Btt|| \quad \ \ (\Att,\Btt \in \mathbfcal{A}) $$
There is the possibility (but not always realizable) that $\mathbfcal{A}$ admits a $\boldsymbol{C^*}$-\textbf{norm}, which is distinguished by the properties
$$ ||\Att^*\Att|| = ||\Att||^2 \ \ \ \ \text{and} \ \ \ \ ||\mathbf{1}|| = 1\,,\ \ \ \text{implying} \ \
||\Att^*|| = ||\Att|| $$
A unital $*$-algebra $\mathbfcal{A}$ which admits a $C^*$-norm and is complete in that 
$C^*$-norm is called a $\boldsymbol{C^*}$-\textbf{algebra}.
The $C^*$-norm of a $C^*$-algebra is unique. 
\\[10pt]
Recall: $\mathbfcal{A}$ is complete with respect to the norm $||\,.\,||$ if every Cauchy sequence
of elements in $\mathbfcal{A}$ w.r.t.\ $||\,.\,||$ converges w.r.t. $||\,.\,||$ to a unique element in $\mathbfcal{A}$. Convergence of a sequence $\{\Att_n\}_{n \in \mathbb{N}}$ of elements in $\mathbfcal{A}$ to an element $\Att\in \mathbfcal{A}$ w.r.t. $||\,.\,||$ means
$$ \lim_{n \to \infty}\, || \Att_n - \Att|| = 0 $$
${}$\\
One reason for considering $C^*$ algebras is that they admit (non-trivial) Hilbert space representations. This brings $C^*$-algebras into close contact with the mathematical framework of quantum mechanics.

For a $C^*$-algebra $\mathbfcal{A}$, a pair $(\pi,\mathcal{H})$ is called a \textbf{(unital) Hilbert space $*$-representation} if
\begin{itemize}
 \item $\mathcal{H}$ is a Hilbert space
 \item $\pi : \mathbfcal{A} \to \mathsf{B}(\mathcal{H})$ is a linear map, with 
 $$ \pi(\Att\Btt) = \pi(\Att)\pi(\Btt)\,, \ \ \ \pi(\mathbf{1})= \mathbf{1}_\mathcal{H}\,, \ \ \ 
 \pi(\Att^*) = \pi(\Att)^* $$
with \ \ $\mathbf{1}_\mathcal{H}\psi = \psi$ $(\psi \in \mathcal{H})$ \ \  and 
$\pi(\Att)^*$ is the adjoint operator of $\pi(\Att)$ in $\mathsf{B}(\mathcal{H})$:
$$ \langle \psi,\pi(\Att)^*\varphi\rangle = \langle \pi(\Att)\psi, \varphi \rangle \quad \ \ (\psi,\varphi \in \mathcal{H})\,, \ \ \text{where} $$
$$ \langle \psi, \varphi \rangle \quad \ \ \text{is the scalar product of} \ \ \psi,\varphi \in \mathcal{H} $$
\end{itemize}
\textbf{Remark}. With the forming of the adjoint operator $A^*$ for every $A \in \mathsf{B}(\mathcal{H})$ as star operation, composition of operators as algebra product, the unit operator 
$\mathbf{1}_\mathcal{H}$ as unit element and the operator norm
$$ || A ||_{op} = \sup\{||A \psi|| : || \psi || = \sqrt{\langle \psi,\psi\rangle} = 1\} $$
as $C^*$-norm, $\mathsf{B}(\mathcal{H})$ is naturally a $C^*$-algebra (equalling its ``identical representation'' on $\mathcal{H}$).
${}$\\[6pt]
Let $\mathbfcal{A}$ be a unital $C^*$-algebra (or just a unital $*$-algebra). 
\\[6pt]
A linear functional $\omega: \mathbfcal{A} \to \mathbb{C}$ is called a
\textbf{state}  if
$$ \omega(\Att^*\Att) \ge 0 \quad (\Att \in \mathbfcal{A}) \quad \ \ \ \text{and} \quad \ \ \ \omega(\textbf{1}) = 1 $$
The first property is called \textbf{positivity}, meaning that any positive element of the algebra, i.e.\ any one that can be written as a ``square'' $\Att^*\Att$, evaluates in a state 
to give a non-negative number. Together with the second property, called \textbf{normalization},
a state can be seen as the generalization of a probability measure: Probabilities are non-negative, and the probability of the ``certain event'' is equal to 1.
\\[10pt]
If $(\pi,\mathcal{H})$ is a unital Hilbert space $*$-rep.\ of a $C^*$-algebra $\mathbfcal{A}$ and 
$\boldsymbol{\varrho}$ is a \textbf{density matrix} on $\mathcal{H}$, then 
$$ \omega_{\pi,\boldsymbol{\varrho}}(\Att) = \text{Tr}(\boldsymbol{\varrho} \pi(\Att)) $$
defines a state on $\mathbfcal{A}$. \\ (Density matrix = positive operator in $\mathsf{B}(\mathcal{H})$ with $\text{Tr}(\boldsymbol{\varrho}) = 1$.)
\\[6pt]
The set
$$\text{Fol}(\pi,\mathcal{H}) = \text{set of all} \ \ \omega_{\pi,\boldsymbol{\varrho}}\,, \ \ \text{where} \ \ \boldsymbol{\varrho} = \text{density matrix on} \ \mathcal{H} $$
is called the \textbf{folium of states} of the rep.\ $(\pi,\mathcal{H})$ for a $C^*$-algebra $\mathbfcal{A}$. 
\\[6pt]
An important theorem assures that \textbf{every state on a $C^*$-algebra lies in the folium of some Hilbert space representation}:
\\[6pt]
\textbf{Theorem 1.1.A (GNS-representation)} \ \ For a state $\omega$ on a unital $C^*$-algebra $\mathbfcal{A}$, there is a (uniquely determined, up
to unitary equivalence) \textbf{GNS triple} $(\pi_\omega,\mathcal{H}_\omega,\Omega_\omega)$ \,:
\begin{itemize}
 \item $(\pi_\omega,\mathcal{H}_\omega)$ is a unital $*$-representation of $\mathbfcal{A}$
 \item $\Omega_\omega$ is a unit vector in $\mathcal{H}_\omega$
 \item $\omega(\Att) = \langle \Omega_\omega,\pi_\omega(\Att)\Omega_\omega \rangle = \text{Tr}(|\Omega_\omega \rangle \langle \Omega_\omega | \pi_\omega(\Att))$
 \ \ $(\Att \in \mathbfcal{A})$
 \item $\Omega_\omega$ is a \textbf{cyclic vector} for $(\pi_\omega,\mathcal{H}_\omega)$, i.e.
 $$ \pi_\omega(\mathbfcal{A})\Omega_\omega = \{\pi_\omega(\Att)\Omega_\omega: \Att \in \mathbfcal{A}\,\} \ \ \text{is dense in} \ \ \mathcal{H}_\omega $$
\end{itemize}
Two unital $*$-rep.s $(\pi,\mathcal{H})$ and $(\tilde{\pi},\tilde{\mathcal{H}})$ of a unital $C^*$-algebra $\mathbfcal{A}$ are \textbf{unitary equivalent}
it there is a unitary operator $U: \mathcal{H} \to \tilde{\mathcal{H}}$ with 
$$ U \pi(\Att) = \tilde{\pi}(\Att)U \quad \ \ \ (\Att \in \mathbfcal{A}) $$
If $\omega$ and $\sigma$ are two states on a unital $C^*$-algebra $\mathbfcal{A}$, and if $(\pi_\omega,\mathcal{H}_\omega)$ and 
$(\pi_\sigma,\mathcal{H}_\sigma)$ are unitary equivalent, then $\text{Fol}(\pi_\omega,\mathcal{H}_\omega) = \text{Fol}(\pi_\sigma,\mathcal{H}_\sigma)$.
\\[6pt]
There is a converse if the rep.s are \textbf{faithful}, meaning that 
$$\pi_\omega(\Att) = 0 \ \ \Longrightarrow \ \ \Att = 0$$
In this case, there is a $C^*$-algebra isomorphism 
$$ \alpha = \pi_\sigma \circ \pi_\omega^{-1} : \pi_\omega(\mathbfcal{A}) \to \pi_\sigma(\mathbfcal{A}) $$
and the question is if that can be extended to a von Neumann algebra isomorphism -- see below.
${}$\\[6pt]
A state on a unital $*$-algebra $\mathbfcal{A}$ is called \textbf{faithful} if, for any $\Att \in \mathbfcal{A}$,
$$ \omega(\Att^*\Att) = 0 \quad \ \ \ \Longrightarrow \quad \ \ \ \Att = 0 $$
For a unital $C^*$-algebra, if a state $\omega$ is faithful, then  $(\pi_\omega,\mathcal{H}_\omega)$ is 
faithful.
\\[6pt]
Any unital $C^*$-algebra $\mathbfcal{A}$ possesses faithful representations. So the algebraic structure of a unital $C^*$-algebra 
can be recovered from its Hilbert space representations.
\\[6pt]
A unital $*$-algebra $\mathbfcal{A}$ need not possess faithful representations and hence need not admit faithful states (in a sense, no other than ``trivial'' states). 
But many unital $*$-algebras which are not $C^*$-algebras are known to admit ``non-trivial'' states, and for those states, one can 
define a generalized concept of ``GNS quadruples''. A difficulty arises since the case may occur that the representers (Hilbert space operators) of the algebra elements might be unbounded, so they are no longer defined as operators on the full Hilbert space, and a dense domain of definition must 
be specified. 
${}$\\[6pt]
Let $\mathbfcal{A}$ be a $*$-algebra with unit ${\bf 1}$, and let $\omega$ be a state on $\mathbfcal{A}$\,.
\\[6pt]
Then there is the \textbf{Wightman-GNS quadruple} $(\mathcal{H}_\omega,\mathcal{D}_\omega,\Pi_\omega,\Omega_\omega)$, (unique up to unitary equivalence) where
\begin{itemize}
 \item $\mathcal{H}_\omega$ is a Hilbert space, \quad $\mathcal{D}_\omega \subset \mathcal{H}_\omega$ dense subspace 
 \item $\Omega_\omega \subset \mathcal{D}_\omega$ is unit vector 
 \item For every $\Att \in \mathbfcal{A}$, $\Pi_\omega(\Att)$ is a linear operator defined on $\mathcal{D}_\omega$, and $\Pi_\omega(\Att)\mathcal{D}_\omega \subset 
 \mathcal{D}_\omega$.
 \item On $\mathcal{D}_\omega$: $\Pi_\omega(c\Att + \Btt) = c\Pi_\omega(\Att) + \Pi_\omega(\Btt)$\,; \ \ \ $\Pi_\omega(\Att\Btt) = \Pi_\omega(\Att)\Pi_\omega(\Btt)$, $\Pi_\omega(\mathbf{1}) = \mathbf{1}_{\mathcal{H}_\omega}$ \ \
 $(c \in \mathbb{C}\,, \ \Att,\Btt \in
 \mathbfcal{A}$)\\[2pt]
 i.e.\ $\Pi_\omega$ is a representation of $\mathbfcal{A}$ by densely defined linear operators in $\mathcal{H}_\omega$
 \item $\Pi_\omega(\Att)^*|_{\mathcal{D}_\omega} = \Pi_\omega(\Att^*)$, where $\Pi_\omega(\Att)^*|_{\mathcal{D}_\omega}$ denotes the
 restriction of the adjoint operator of $\Pi_\omega(\Att)$ to the domain $\mathcal{D}_\omega$.
 \item $ \omega(\Att) = \langle \Omega_\omega, \Pi_\omega(\Att) \Omega_\omega \rangle_{\mathcal{H}_\omega}$ \quad $(\Att \in \mathbfcal{A})$
 \item $\mathcal{D}_\omega = \Pi_\omega(\mathbfcal{A})\Omega_\omega$ \quad \ (minimal choice)
\end{itemize}
Let $\mathcal{H}$ a Hilbert space.
\begin{itemize}
 \item For $\mathsf{N} \subset \mathsf{B}(\mathcal{H})$, define the \textbf{commutant}
 $$ \mathsf{N}' = \{ A \in \mathsf{B}(\mathcal{H}): AB = BA \ \ \text{for all} \ B \in \mathsf{N}\}$$
 \item $\mathsf{N} \subset \mathsf{B}(\mathcal{H})$ is called a \textbf{von Neumann algebra} if 
 $\mathsf{N}$ equals its \textbf{bi-commutant}, i.e.
 $$ \mathsf{N} = (\mathsf{N}')' $$
 \item $\mathsf{N} \subset \mathsf{B}(\mathcal{H})$ is a von Neumann algebra if and only if $\mathsf{N}$ 
 is a $*$-algebra under the natural operations of forming linear combinations, products and adjoints 
 in $\mathsf{B}(\mathcal{H})$, if $\mathsf{N}$ contains the unit operator $\mathbf{1}_\mathcal{H}$, and if 
 $\mathsf{N}$ is \textbf{closed in the weak operator topology}: This means, if 
 $$ A_n \to A \ \text{weakly} \ \ \Longleftrightarrow \langle \psi, (A_n - A) \chi \rangle \to 0 \ \ (\psi,\,\chi \in \mathcal{H}) $$
 for any sequence $A_n \in \mathsf{N}$ with some $A \in \sf{B}(\mathcal{H})$, then $A \in \mathsf{N}$. 
\end{itemize}
\textbf{Exercise} \ \ Show that every vN algebra is also a sub-$C^*$-algebra of $\mathsf{B}(\mathcal{H})$, but not vice versa.
${}$\\[6pt]
If $\mathbfcal{A}$ is a unital $C^*$-algebra, and $\pi : \mathbfcal{A} \to \mathsf{B}(\mathcal{H})$ is a unital $*$-representation,
then
$$ \mathsf{N} = \mathsf{N}_\pi = \pi(\mathbfcal{A})'' $$
is called the \textbf{induced vN algebra of} $\mathbfcal{A}$ \textbf{in the rep.} $(\pi,\mathcal{H})$.
\\[6pt]
Properties of the induced vN algebra relate to properties of the representation.
We define:
\\[6pt] 
A state $\omega$ on $\mathbfcal{A}$ is called \textbf{pure} if it is \underline{not} a (non-trivial) convex sum of different states. A convex sum of states $\omega_1,\ldots,\omega_n$ with weights $\lambda_1,\ldots,\lambda_n$, where $\lambda_j \ge 0$ and 
$\sum \lambda_j = 1$, is the state 
$$ \overline{\omega}(\Att) = \sum_{j = 1}^n \lambda_j \omega_j(\Att) \quad \ \ \ (\Att \in \mathbfcal{A}) $$
\textbf{Theorem 1.1.B} 
\begin{itemize}
 \item A state $\omega$ on a unital $C^*$-algebra $\mathbfcal{A}$ \textbf{is pure if and only if} $(\pi_\omega,\mathcal{H}_\omega)$ \textbf{is irreducible},
 where a rep.\ $(\pi,\mathcal{H})$ of $\mathbfcal{A}$ is called \textbf{irreducible} if 
 $$ \mathsf{N}_\pi = \mathsf{B}(\mathcal{H}) \quad \Longleftrightarrow \quad \mathsf{N}_\pi{}' = \mathbb{C} \mathbf{1}_\mathcal{H} $$
 \item If $\omega$ is pure, then for every unit vector $\psi \in \mathcal{H}_\omega$, the induced state 
 $$\omega_{|\psi\rangle \langle \psi|}(\Att) = \langle \psi, \pi_\omega(\Att)\psi \rangle \quad \ \ 
 \text{on} \ \ \mathbfcal{A} \ \ \text{is pure} $$ 
 \item If $(\pi,\mathcal{H})$ and $(\tilde{\pi},\tilde{\mathcal{H}})$ are faithful and irreducible rep.s of $\mathbfcal{A}$, then 
 they are unitary equivalent if and only if
 $$ \text{Fol}(\pi,\mathcal{H}) = \text{Fol}(\tilde{\pi},\tilde{\mathcal{H}}) $$
\end{itemize}
There is a generalization if reps. $(\pi,\mathcal{H})$ and $(\tilde{\pi},\tilde{\mathcal{H}})$ are not pure, but still faithful:
In this case one says that the reps. are \textbf{quasiequivalent} if the $C^*$-algebra isomorphism 
$$\alpha = \pi_\sigma \circ \pi_\omega^{-1} : \pi_\omega(\mathbfcal{A}) \to \pi_\sigma(\mathbfcal{A}) $$
extends to a vN algebra isomorphism
$$ u: \mathsf{N}_\pi \to \mathsf{N}_{\tilde{\pi}} $$
by taking limits in the weak operator topology.
\\[6pt]
\textbf{Theorem 1.1.C} \ \ Two faithful unital $*$-reps. $(\pi,\mathcal{H})$ and $(\tilde{\pi},\tilde{\mathcal{H}})$ of a unital $C^*$-algebra
\textbf{are quasiequivalent if and only if}
$$ \text{Fol}(\pi,\mathcal{H}) = \text{Fol}(\tilde{\pi},\tilde{\mathcal{H}}) $$
The concept of unitary equivalence or quasiequivalence is important. On a $C^*$-algebra or $*$-algebra containing the observables of $\infty$-many degrees
of freedom there are typically too many states, i.e.\ states which cannot be interpreted as describing physically realistic situations. Any set of states selected according to a criterion of ``physicality'' should define (and hence, lie in) the same, single folium of states, at least ``locally''. 
\\[6pt]
\textbf{Exercise}
\\[2pt]
Consider the Hilbert space $\mathcal{H} = L^2(\mathbb{R},dx)$. Then $\mathsf{B}(\mathcal{H})$ is a $C^*$-algebra. 
One can define states $\overset{\nsim}{\omega}$ on $\mathsf{B}(\mathcal{H})$ with the property that $\overset{\nsim}{\omega}(P) = 0$ whenever 
$P$ is a finite-dimensional projector. 
Show that $\overset{\nsim}{\omega}$ is not induced by a density matrix, i.e.\ there is no density matrix operator $\boldsymbol{\varrho}$ in
$\mathsf{B}(\mathcal{H})$ such that $\overset{\nsim}{\omega}(A) = \mathrm{Tr}(\boldsymbol{\varrho} A)$ holds for all $A \in \mathsf{B}(\mathcal{H})$. Conclude that $\overset{\nsim}{\omega}$ is not contained in the folium of the GNS representation of any state $\omega_\psi(A) = \langle \psi,A\psi \rangle$ for $\psi$ a unit vector in $\mathcal{H}$. 
\\[6pt]
In Tolksdorf and Verch (2018), such a state $\overset{\nsim}{\omega}$ is constructed as a limit of thermal equilibrium states for the 1-dimensional harmonic oscillator as the temperature diverges. Therefore, such a state is also referred to as the ``hell'' state.

\subsection{Algebraic QFT on Minkowski Spacetime: Haag-Kastler Nets of Local Algebras}

Let $\mathcal{M} \simeq \mathbb{R}^{1,3}$ be Minkowski spacetime, with spacetime metric $\eta = (\eta_{\mu \nu})_{\mu,\nu = 0}^3 = (1,-1,-1,-1)$ w.r.t. inertial coordinates. (The following concept of $J^\pm(p)$ and of ``double cone'' can naturally be generalized to generic curved spacetimes.) We define: 
 $J^\pm(p)  = $ \  set of all $q \in \mathcal{M}$ on any future(+)/past(--) directed worldline emanating from  $p \in \mathcal{M}$, and 
 $J^\pm(S) =   \bigcup_{p \in S} J^\pm(p)$ for $S \subset \mathcal{M}$. Then
 $O =$\ open interior of $J^+(p) \cap J^-(p')$ for $p' \in J^+(p)$ is called a ``double cone''. 
 \begin{center}
 \includegraphics[width=16.3cm]{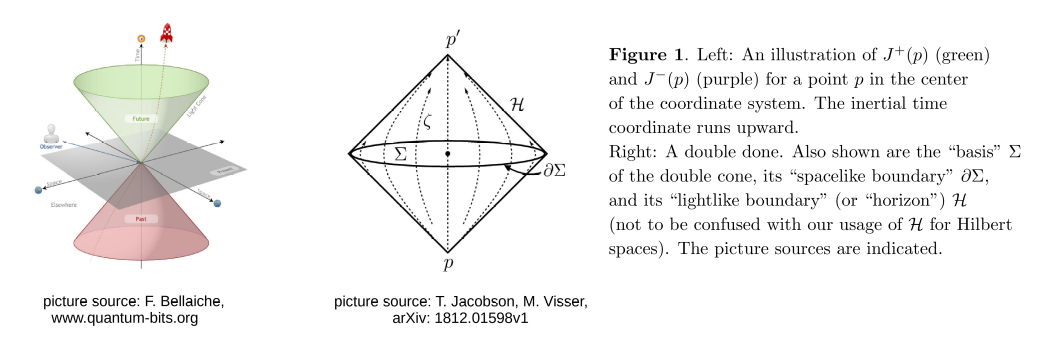}
 \end{center}
 \newpage \noindent
In algebraic quantum field theory, there is 
a local structure for the algebra $\mathbfcal{A}$ of observables of a quantum field
\begin{align*}
 &\mathbfcal{A}  = \ \text{*-algebra of (or: generated by) observables}, \\[4pt] & \quad \text{formed by *-subalgebras}\\
 &\mathbfcal{A}(O) = \ \text{algebra of observables that can be measured in the spacetime region} \ { O}
\end{align*}
with the properties:
\begin{itemize}
 \item $O_1 \subset O_2$ \quad $\Longrightarrow$ \quad $\mathbfcal{A}(O_1) \subset \mathbfcal{A}(O_2)$ 
 \item $O_2 \cap J^\pm(O_1) = \emptyset$ \quad $\Longrightarrow$ \quad $[\Att_1,\Att_2] = 0$ for all $\Att_j \in \mathbfcal{A}(O_j)$ \quad \ \ $([\Att_1,\Att_2] = \Att_1\Att_2 - \Att_2\Att_1)$
 \item For every symmetry (isometry) $L : M \to M$ of the spacetime, there is an automorphism $\alpha_L: \mathbfcal{A} \to \mathbfcal{A}$
 so that 
 $$ \alpha_L(\mathbfcal{A}(O)) = \mathbfcal{A}(L(O)) \quad \text{and} \quad \alpha_{L_1} \circ \alpha_{L_2} = \alpha_{L_1 L_2} $$
\end{itemize}
The algebras $\mathbfcal{A}(O)$ are non-commutative for a quantum field theory. The collection of all $\mathbfcal{A}(O)$, or rather 
the assignment $O \to \mathbfcal{A}(O)$ of local spacetime regions to unital $*$-algebras, is frequently called a \textbf{local net, or Haag-Kastler-net, of observable algebras}. 
\\[6pt]
Assume that $\mathbfcal{A}$ is a unital $C^*$-algebra and that the $\mathbfcal{A}(O)$ are unital sub-$C^*$-algebras (for simplicity).
Let $\mathcal{G} \subset \text{Iso}(\mathcal{M})$ be a subgroup of the spacetime isometry group.
\\[6pt]
A state $\omega$ on $\mathbfcal{A}$ is called $\mathcal{G}$-\textbf{invariant} if
$$ \omega \circ \alpha_L = \omega \quad \ \ \ (L \in \mathcal{G})$$
If the state $\omega$ is $\mathcal{G}$-invariant and $(\pi_\omega,\mathcal{H}_\omega,\Omega_\omega)$ is the GNS-triple,
then there is a unitary group representation $\{U_L\}_{L \in \mathcal{G}}$ on $\mathcal{H}_\omega$ with the properties
$$ U_L\Omega_\omega = \Omega_\omega \quad \ \ \text{and} \ \ \quad \pi_\omega(\alpha_L(\Att)) = U_L \pi_\omega(\Att) U_L^* \ \ \quad (L \in \mathcal{G}\,, \ \Att \in \mathbfcal{A}) $$
One then say that the unitary group $\{U_L\}_{L \in \mathcal{G}}$ implements the automorphism group $\{\alpha_L\}_{L \in \mathcal{G}}$
in the GNS representation $(\pi_\omega,\mathcal{H}_\omega,\Omega_\omega)$ of $\omega$.
\\[6pt]
\textbf{Exercise} \ \ Prove this. Discuss also what properties of $\alpha_L$, expressed at the level of $\omega$ and the elements of $\mathbfcal{A}$, are 
sufficient to conclude that $U_L$ is strongly continuous w.r.t. $L$, in case that $\mathcal{G}$ is a Lie group.
\\[6pt]
For Minkowski spacetime, we consider $\mathcal{P}^\uparrow_+$, the \textbf{proper, orthochronous Poincar\'e group} as the spacetime symmetry group.
\\[6pt]
For any timelike, future directed vector $v$, there is a one-parametric subgroup $\{ L^{(v)}_t\}_{t \in \mathbb{R}}$ of $\mathcal{P}^\uparrow_+$:
$$ L^{(v)}_t x = x + tv \quad \ \ (x \in \mathbb{R}^{1,3}) \quad \ \text{ ``time-shifts in direction}\ \, v {}\text{''} $$
On writing $\alpha^{(v)}_t = \alpha_{L_t^{(v)}}$,
\begin{itemize}
 \item A state $\omega$ on $\mathbfcal{A}$ is a \textbf{ground state} for $\{\alpha^{(v)}_t\}_{t \in \mathbb{R}}$ if  
 $\omega \circ \alpha_t^{(v)} = \omega$ and 
 $$ \int_{\mathbb{R}} f(t) \,\omega(\Att \alpha_t^{(v)}(\Btt)) \, dt = 0 \ \ \quad (\Att,\Btt \in \mathbfcal{A}) $$
 for all Schwartz-type test-functions $f$ so that $\text{supp}(\widehat{f}) \subset (-\infty,0)$, where
 $$ \widehat{f}(k) = \frac{1}{\sqrt{2\pi}}\int_{\mathbb{R}} f(t) \,e^{-ikt}\,dt \ \ \ \ \ \text{(Fourier transform)} $$
\end{itemize}
\begin{itemize}
 \item A state $\omega$ on $\mathbfcal{A}$ is a \textbf{vacuum state} if $\omega \circ \alpha_L = \omega$ for all $L \in \mathcal{P}^\uparrow_+$ and 
 if $\omega$ is a ground state for $\{\alpha_t^{(v)}\}_{t \in \mathbb{R}}$ for any (and hence, for every) timelike, future directed Minkowski vector
 $v$. 
\end{itemize}
Implicit assumption here: The unitary group $\{U_L\}_{L \in \mathcal{P}^\uparrow_+}$ which implements $\{\alpha_L\}_{L \in \mathcal{P}^\uparrow_+}$
in the GNS-triple $(\pi_\omega,\mathcal{H}_\omega,\Omega_\omega)$ is continuous. 
\\[6pt]
\textbf{Exercise} \ \ If $\omega$ is a ground state, show that the one-parametric unitary group $\{U_t^{(v)}\}_{t \in \mathbb{R}}$ implementing 
$\{\alpha_t^{(v)}\}_{t \in \mathbb{R}}$ in the GNS triple $(\pi_\omega,\mathcal{H}_\omega,\Omega_\omega)$ has a non-negative generator, i.e.
$U^{(v)}_t = e^{itH^{(v)}}$ with a ``Hamilton operator'' $H^{(v)} \ge 0$
\\[6pt]
In QFT on Minkowski spacetime, \textbf{the vacuum state is viewed as a physical state}. In keeping with previous discussion, one expects
$$ \text{Fol}(\pi_\omega) \subset \{\,\text{all ``physical states''}\,\} \quad \ \ (\omega = \ \text{vacuum state})$$
But: Don't lightly put ``$=$''here instead of ``$\subset$'' --- doing so might rule out different phases or charges.
\\[6pt]
Once a physical state is chosen --- in this case, the vacuum state $\omega$ --- one can form the \textbf{local von Neumann algebras}
$$ \mathsf{N}_\omega(O) = \pi_\omega(\mathbfcal{A}(O))'' $$
They have the properties 
\begin{itemize}
 \item $O_1 \subset O_1$ \ \ $\Longrightarrow$ \ \ $\mathsf{N}_\omega(O_1) \subset \mathsf{N}_\omega(O_2)$
 \item $O_2 \cap J^\pm(O_1) = \emptyset$ \quad $\Longrightarrow$ \quad $[A_1,A_2]= 0$ for all $A_j \in \mathsf{N}_\omega(O_j)$
 \item $U_L \mathsf{N}_\omega(O) U_L^* = \mathsf{N}_\omega(L(O))$ \quad $(L \in \mathcal{P}^\uparrow_+)$ \\
 with the unitary group $\{U_L\}_{L \in \mathcal{P}^\uparrow_+}$ implementing $\{\alpha_L\}_{L \in \mathcal{P}^\uparrow_+}$
 in the GNS triple of the vacuum state $\omega$. 
\end{itemize}
\textbf{Exercise} \ \ Prove this.
\\[10pt]
Passing to the local von Neumann algebras has a major advantage: \\[6pt]
Any $A = A^* \in \mathsf{N}_\omega(O)$ is interpreted as a [mathematical expression for a potential] \textbf{observable}.
\\[6pt]
The statistical interpretation of quantum theory is provided by the \textbf{spectral measure}, or \textbf{projector valued measure},
$\{P^{(A)}(J)\}_{J \subset \mathbb{R}}$ that is associated with every self-adjoint $A$, by 
\begin{align*}
\langle P^{(A)}(J) \rangle_{\boldsymbol{\varrho}} & = \text{Tr}(\boldsymbol{\varrho} P^{(A)}(J)) \\
 & = \text{probability of finding values in the set} \ J \ \text{on measuring} \\
 & \text{the observable} \ A  
 \ \text{in a state 
 corresponding to the density matrix} \ \boldsymbol{\varrho}
\end{align*}
For vN algebras we have:
$$ A = A^* \in \mathsf{N}_\omega(O) \quad \Longleftrightarrow  \quad P^{(A)}(J) \in \mathsf{N}_\omega(O) \ \ \ \ (J \ \text{any Borel subset of} \ \mathbb{R}) $$
\textbf{Homework} \ \ Recapitulate the spectral theorem for self-adjoint operators and its significance for the statistical/probabilistic interpretation 
of quantum mechanics.
\\ \\
{\small
\textbf{References}\\[2pt]
The material on operator algebras and the Haag-Kastler approach of localizable observables in terms of local nets of operator algebras 
on Minkowski spacetime presented in this section is standard textbook material, and taken from the following references:
\\[6pt]
Araki, H., \textit{Mathematical Theory of Quantum Fields}, Oxford University Press, 1999
\\[4pt]
Bratteli, O., Robinson, D.W., \textit{Operator Algebras and Quantum Statistical Mechanics}, Vols.\ 1 and 2, Springer-Verlag, 1987/1997
\\[4pt]
Haag, R., \textit{Local Quantum Physics}, Springer-Verlag, 1996
\\[4pt]
Pedersen, G.K., \textit{ $C^*$-Algebras and Their Automorphism Groups}, Academic Press, 1979
\\[4pt]
Tolksdorf, J., Verch, R. (2018), ``Quantum physics, fields and closed timelike curves: The D-CTC condition in quantum field theory'',
Commun.\ Math.\ Phys.\ \textbf{357}, 319–351
\\[10pt]
\textbf{Further Reading}
\\[6pt]
Baumg\"artel, H., Wollenberg, M., \textit{Causal Nets of Operator Algebras: Mathematical aspects of algebraic quantum field theory},
Akademie-Verlag, 1992
\\[4pt]
Brunetti, R., Dappiaggi, C., Fredenhagen, K., Yngvason, J. (eds.), \textit{Advances in Algebraic Quantum Field Theory}, Springer-Verlag, 2016
\\[4pt]
Streater, R., Wightman, A.S., \textit{PCT, Spin and Statistics, and All That}, Princeton University Press, 2000 (revised edition)
\\[4pt]
Sunder, V.S., \textit{An Invitation to von Neumann Algebras}, Springer-Verlag, 1987
\\[4pt]
Haag, R., Kastler, D. (1964), ``An algebraic approach to quantum field theory'', J.\ Math.\ Phys.\ \textbf{5}, 848-861
\\[4pt]
Halvorson, H., M\"uger, M. (2007), ``Algebraic quantum field theory'', in: J.\ Butterfield and
J.\ Earman (Eds.), \textit{Handbook of the Philosophy of Science}, Elsevier, pp.\ 731–922
}
\newpage \noindent
\section{Tomita-Takesaki Modular Theory of von Neumann algebras and the Bisognano-Wichmann/Borchers Theorem}

\subsection{Tomita-Takesaki Modular Theory}

This section begins by collecting some definitions and facts about closable and closed operators.
\begin{itemize}
 \item 
 Let $\mathcal{H}$ be a Hilbert space, \quad $\mathcal{D} \subset \mathcal{H}$ dense linear subspace, 
 \item 
  $ T : \mathcal{D} \to \mathcal{H}$ an anti/linear operator \\[2pt]
  (antilinear operator is like linear operator, except for $T(a\psi) = \overline{a}T(\psi)$ \ \ for all 
 $a \in \mathbb{C}$, $\psi \in \mathcal{D})$ 
\item The \textbf{graph of} $T$ is the set $\Gamma(T) = \{\psi \oplus T\psi: \psi \in \mathcal{D}\} \subset \mathcal{D} \oplus \mathcal{H}$ 
\item The \textbf{graph norm} is \ \ $||\psi \oplus T \psi||_{\Gamma(T)} = (|| \psi ||^2 + || T\psi ||^2)^{1/2}$ 
with $||\,.\,||$ the norm on $\mathcal{H}$ 
\item $(T,\mathcal{D})$ is \textbf{closable} if $\overline{\Gamma(T)}$, the closure of $\Gamma(T)$ in 
$\mathcal{H} \oplus \mathcal{H}$, is the graph of a (unique) operator, called the 
\textbf{closure} of $(T,\mathcal{D})$, denoted by $(T_c,\mathcal{D}_c)$. 
$\xi \oplus T_c\xi \in \Gamma(T_c) = \overline{\Gamma(T)}$ if and only if there is a sequence $\{\psi_n\}_{n \in \mathbb{N}}$, $\psi_n \in \mathcal{D}$ with 
$$ || \psi_n - \xi || + || T\psi_n - T_c\xi || \to 0 \quad (n \to \infty) $$
\item $(T,\mathcal{D})$ is \textbf{closed} if $(T,\mathcal{D}) = (T_c,\mathcal{D}_c)$. 
 \item A bounded operator $(T,\mathcal{H})$ defined on the full Hilbert space is closed. 
 \item An operator $(T,\mathcal{D})$ is closable if and only if the adjoint operator is densely defined. 
 Consequently: Symmetric operators and bounded operators are closable. Counterexamples to closable operators
 occur only for unbounded operators (in $\infty$-dim.\ Hilbert space).
 \end{itemize}
\textbf{Theorem 2.1.A: Polar decomposition of closed operators} \\[4pt]
Let $(T,\mathcal{D})$ be a closed operator. Then $T$ can be \textit{uniquely} written in the form 
$$ T = V|T| $$
where:
\begin{itemize}
 \item[(1)] $V$ is an \textit{anti/linear partial isometry}, i.e.\ a bounded anti/linear operator with the 
 properties 
 $$ V^*V = \text{projector on} \ \overline{\text{Range}(|T|)}\,, \quad \ \ VV^* = \text{projector on} \  \overline{\text{Range}(T)} $$
 \item[(2)]
 $|T|$ is a \textit{positive, selfadjoint operator} defined on $\mathcal{D}$, i.e.\ $\langle \psi,|T|\psi \rangle \ge 0$
 for all $\psi \in \mathcal{D}$; moreover, 
 $$ \langle |T|\psi,|T|\psi\rangle = \langle T\psi,T\psi \rangle \ \quad \ (\psi \in \mathcal{D}) $$
\end{itemize}
${}$\\[4pt]
 Let $\mathsf{N} \subset \mathsf{B}(\mathcal{H})$ be a von Neumann algebra, $\psi \in \mathcal{H}$ a unit vector. 
\begin{itemize}
 \item $\psi$ is called \textbf{cyclic} for $\mathsf{N}$ if 
 $$ \mathsf{N}\psi = \{A\psi: A \in \mathsf{N}\} \quad \text{is dense in} \ \mathcal{H}$$
 \item $\psi$ is called \textbf{separating} for $\mathsf{N}$ if $A\psi = 0 \Rightarrow A = 0$ for all $A\in \mathsf{N}$
 \item $\psi$ is called a \textbf{standard vector} for $\mathsf{N}$ if $\psi$ is \textbf{cyclic and separating} for $\mathsf{N}$. 
\end{itemize}
The following statement provides a nice exercise:
$$ \mathsf{N} \ \text{is von Neumann algebra} \quad \Longrightarrow \quad \mathsf{N}' \ \text{is von Neumann algebra} $$
$$ \psi \ \text{is cyclic\,/\,separating for} \ \mathsf{N} \quad \Longleftrightarrow \quad 
\psi \ \text{is separating\,/\,cyclic for} \ \mathsf{N}'$$
Let $\mathsf{N} \subset \mathsf{B}(\mathcal{H})$ be a von Neumann algebra, $\psi \in \mathcal{H}$ standard vector
\begin{itemize}
 \item Then $\mathcal{D} = \mathcal{D}_{\mathsf{N},\psi}$ defined by $\mathcal{D} = \mathsf{N}\psi$ is a dense subspace of $\mathcal{H}$, since $\psi$ is cyclic for $\mathsf{N}$.
 \item An anti-linear operator $(S,\mathcal{D})$ can be defined by setting 
 $$ S(A\psi) = A^*\psi  \quad \ \ (A \in \mathsf{N}) \quad \quad \textbf{Tomita operator} $$
 Note that this definition leads to a well-defined map from $\mathcal{D}$ to $\mathcal{D}$ since $\psi$ is separating for $\mathsf{N}$. This means that $A$ is uniquely defined by $A\psi$. If there is potentially another $\tilde{A} \in \mathsf{N}$ with $\tilde{A}\psi = A\psi$ $\Longrightarrow$ $(\tilde{A} - A)\psi = 0$ $\Longrightarrow$ $\tilde{A} - A = 0$. Therefore, $S(A\psi) = A^*\psi$ is uniquely defined.
 \item
 Let $A \in \mathsf{N}$, $B \in \mathsf{N}'$. Then $B^*A^* = A^*B^*$, hence
 $$ |\langle B\psi,S(A\psi)\rangle| = |\langle B\psi,A^*\psi \rangle| = |\langle A\psi ,B^*\psi \rangle| \le ||A\psi|| \cdot ||B^*\psi|| $$
 showing that $B\psi$ is in the domain of $S^*$, the adjoint operator of $S$ $\Longrightarrow$ $S^*$ is defined on a dense domain $\Longrightarrow$ $(S,\mathcal{D})$ is closable. 
\end{itemize}
The closure $(S_c,\mathcal{D}_c)$ of the Tomita operator $(S,\mathcal{D})$ defined for $\mathsf{N}$ and 
$\psi$ possesses a polar decomposition, denoted by 
$$ S_c = J \Delta^{1/2} $$
Here: $J$ is anti-unitary, and $\Delta^{1/2} = |S_c|$ is positive and selfadjoint on $\mathcal{D}_c$
\\[6pt]
$J$ is called the \textbf{modular conjugation} and \\[2pt] $\Delta = S_c^*S_c$ is called the \textbf{modular operator} of $\mathsf{N}$ and $\psi$. 
\\[6pt]
Define:
\begin{itemize}
 \item $j(A) = JAJ^*$. Since $J$ is anti-unitary, $j: \mathsf{B}(\mathcal{H}) \to \mathsf{B}(\mathcal{H})$ is an antilinear automorphism. 
 \item $\Delta^{it} = e^{i \ln(\Delta) t}$ $(t \in \mathbb{R})$. This defines a unitary group on 
 $\mathcal{H}$ $\Longrightarrow$ $\sigma_t(A) = \Delta^{it} A \Delta^{-it}$ $(t \in \mathbb{R},\,\,
 A \in \mathsf{B}(\mathcal{H})$) is a one-parametric group of automorphisms of $\mathsf{B}(\mathcal{H})$. $\Delta^{it}$ $(t \in \mathbb{R}$) is called \textbf{modular group}.
\end{itemize}
The modular objects have some interesting (and not entirely obvious) properties which goes to show that they are 
very intimately related with the structure of a von Neumann algebra.
\\[10pt]
\textbf{Theorem 2.1.B: ``Tomita-Takesaki Theorem''} (Takesaki (1970))\\[4pt]
For $\mathsf{N} \subset \mathsf{B}(\mathcal{H})$ a vN algebra with standard vector $\psi \in \mathcal{H}$ the associated \textbf{modular objects} $J$ and $\Delta$ have the following properties:
\begin{itemize}
 \item[(1)] $J = J^*$, \ \ $J^2 = \mathbf{1}$, \ \ $J\psi = \psi$ 
 \item[(2)] $j(\mathsf{N}) = \mathsf{N}'$\,, \ \ $\Delta^{-1/2} = J \Delta^{1/2} J$\,, \ \ $\Delta^{1/2}\psi = \psi$ 
 \item[(3)] $\sigma_t$ is an automorphism of $\mathsf{N}$ for every $t \in \mathbb{R}$; \ \
 $\sigma_t(\mathsf{N}) = \mathsf{N}$
 \item[(4)] Denote by $\omega_\psi(A) = \langle \psi,A\psi \rangle$ $(A \in \mathsf{N})$ the state induced by the standard vector $\psi$ on $\mathsf{N}$. This state is a \textbf{KMS state for the automorphism group $\sigma_t$ $(t \in \mathbb{R})$ at inverse temperature} $\beta = 1$:
 For every $A,B \in \mathsf{N}$ there is an analytic function 
 $$ F_{AB}^{(\beta)}: \{z = t + is \in \mathbb{C}: 0 < s < \beta, \ t \in \mathbb{R}\} \to \mathbb{C}\,, \quad \quad \text{with} $$
 $$ \lim_{s \to 0}\, F_{AB}^{(\beta)}(t + is) = \omega_{\psi}(A\sigma_t(B))\,, \quad 
 \lim_{s \to \beta}\, F_{AB}^{(\beta)}(t + is) = \omega_{\psi}(\sigma_t(B)A) $$
  \item[(5)] Let $V_t$ $(t \in \mathbb{R})$ be another unitary group, 
  \\[2pt]
  $\tau_t(A) = V_t A V_t^*$, \ \ assume that $\tau_t(\mathsf{N}) = \mathsf{N}$.
  \\[2pt]
  If $\omega_\psi$ is a KMS state for the automorphism group $\tau_t$ $(t \in \mathbb{R})$ at 
  inverse temperature $\beta = 1$, then $V_t = \Delta^{it}$ $(t \in \mathbb{R})$.
\end{itemize}
\textbf{Exercise} \ \ (a) Show that the KMS condition at $\beta = 1$ amounts to $\langle \Delta^{1/2}A^*\psi,\Delta^{1/2}B\psi\rangle = \langle B^*\psi,A\psi \rangle$ for $A,B \in \mathsf{N}$. Show that this relation follows from the definition of the modular objects. 
\\[4pt]
(b) Show that the KMS condition is equivalent to the following condition: 
\begin{align*}
 \int_\mathbb{R} \chi(t) \omega(A_1 \sigma_t(A_2)) \,dt = 
  \int_{\mathbb{R}} \chi(t +i\beta) \omega(\sigma_t(A_2)A_1)\,dt \quad \ \
  (A_1,A_2 \in \mathsf{N})
\end{align*}
for all test functions $\chi$ with $\hat{\chi} \in C_0^\infty(\mathbb{R})$.
(Recall that if $\hat{\chi} \in C_0^\infty(\mathbb{R})$, then $\chi$ is a Schwartz type test function which possesses an analytic extension to the complex plane, by the Paley-Wiener theorem.) Hence show that in the limit $\beta \to \infty$, KMS states approximate ground states for 
an automorphism group $\{\sigma_t\}_{t \in \mathbb{R}}$ (need not be the one induced by the modular group).

\subsection{The Bisognano-Wichmann Theorem}
Let $f \mapsto \phi(f)$ $(f \in \mathrm{S}(\mathbb{R}^4))$ be a \textbf{Wightman-type scalar quantum field on Minkowski spacetime} (see Streater and Wightman (2000), loc.\ cit.), with
\begin{itemize}
 \item Hilbert space $\mathcal{H}$, and dense, invariant domain $\mathcal{D}_\phi \subset \mathcal{H}$
 for the field operators $\phi(f)$
 \item continuous unitary representation $U(a,\Lambda)$ $(\,(a,\Lambda) \in \mathcal{P}^\uparrow_+\,)$
 \item vacuum vector $\Omega \in \mathcal{D}_\phi$
\end{itemize}
Assume that the quantum field is \textbf{hermitean}: $\phi(\bar{f}) = \phi(f)^*|_{\mathcal{D}_\phi}$.
\\[2pt]
$\Longrightarrow$ \ \ the $\phi(f)$ are closable. \\ The closures $\phi(f)_c$ have polar decompositions
$$ \phi(f)_c = V_f |\phi(f)_c|     \quad \ \ (f \in \mathrm{S}(\mathbb{R}^4)) $$
For $O$ = open subset of $\mathbb{R}^4$, define:
$$ \mathsf{N}(O) = \{ V_f, b(|\phi(f)_c|) : \mathrm{supp}(f) \subset O, \ b = \text{bounded function} \}'' $$
$\{\mathsf{N}(O)\}_{O \subset \mathbb{R}^4}$ is a \textbf{net of local vN algebras (of observables)}, with the following properties:

\begin{itemize}
 \item \textbf{isotony}: $O_1 \subset O_2 \ \ \Longrightarrow \ \ \mathsf{N}(O_1) \subset \mathsf{N}(O_2) $
 \item \textbf{locality}: $[A_1,A_2 ] = 0$ for $A_j \in \mathsf{N}(O_j)$ if 
 $O_1$ and $O_2$ are \textit{causally separated} 
 \\[2pt]
 \textit{(strictly speaking, this need not follow from locality (spacelike commutativity) of the field 
 operators, and is an additional assumption; however, it is fulfilled in many quantum field model theories that can be rigorously constructed)}
 \item \textbf{covariance}: $U(a,\Lambda) \mathsf{N}(O) U(a,\Lambda)^* = \mathsf{N}(\Lambda O + a)$
 \item \textbf{existence of unique vacuum vector}: $\Omega$ unit vector in $\mathcal{H}$, with 
 $U(a,\Lambda)\Omega = \Omega$, unique up to phase
 \item \textbf{spectrum condition}: If $e\ne 0$ is any causal, future directed vector, then $U(te,1) = e^{itH(e)}$ with $\textrm{spec}(H(e)) \subset [0,\infty)$. 
 \item \textbf{additivity}: If $\{O_\alpha\}_{\alpha \in \mathbb{A}}$ is any covering of $O$, then 
 $$ \mathsf{N}(O) \subset \{ \mathsf{N}(O_\alpha): \alpha\in \mathbb{A} \}'' $$
 \item \textbf{irreducibility}:  $\mathsf{N}(\mathbb{R}^4) = \mathsf{B}(\mathcal{H})$
 \item \textbf{Reeh-Schlieder property}: $\{A\Omega : A \in \mathsf{N}(O)\}$ is dense in $\mathcal{H}$ for any open (non-void) $O \subset \mathbb{R}^4$
 $\Rightarrow$ $\Omega$ is standard vector for any $\mathsf{N}(O)$ (if $O$ has non-void causal complement). 
\end{itemize}
${}$\\
\texttt{The ``Right Rindler Wedge''}
$$ W_R = \{ (x^0,x^1,x^2,x^3) \in \mathbb{R}^4: x^1 > 0,\ |x^0|< x^1 \}\,, \quad \ \  W_L = -W_R = \text{causal complement} $$
\begin{center}
 \includegraphics[width=15.5cm]{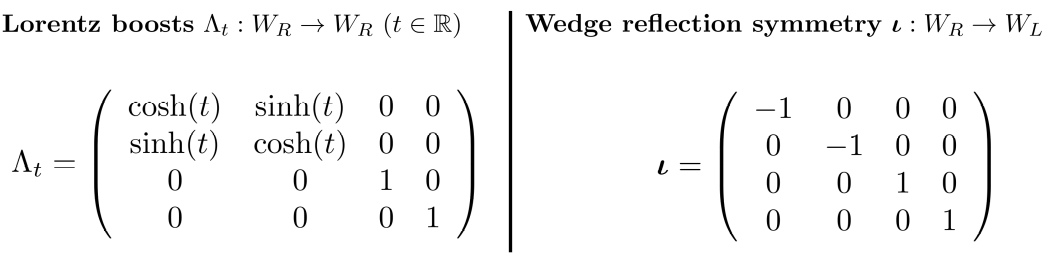}
\end{center}
The following figure provides a spacetime picture of $W_L$ and $W_R$:
 \begin{center}
\includegraphics[width=5.5cm]{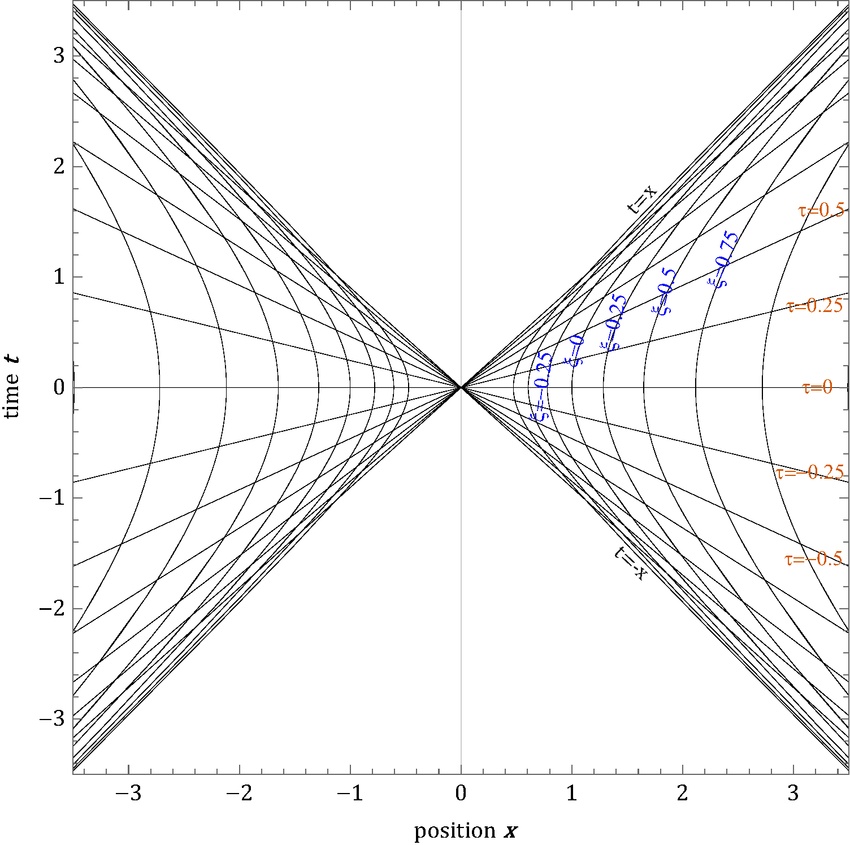}
\end{center}
{\small \textbf{Figure 2.}
The above figure shows $W_L$ and $W_R$. The $t$ in the figure corresponds to our $x^0$. The $\tau$ in the figure corresponds to our $t$. The $x$ in the 
figure corresponds to our $x^1$. The coordinates $x^2$ and $x^3$ are suppressed (would be perpendicular to  the depicted $x^0$-$x^1$ plane).\\[4pt]
The hyperbolae in the wedge regions are orbits of the action of the $\Lambda_t$ $(t \in \mathbb{R})$, i.e.\ a hyperbola corresponds to $ \{ \Lambda_t q : t \in \mathbb{R} \} $
for any $q$ in $W_R$ or $W_L$.\ \ The wedge reflection $\boldsymbol{\iota}$ acts as reflection along the straight lines. 
\\[2pt]
The figure is taken from S.\ Ariwahjoedi et al., arXiv:2309.08833}                     
\\ \\
\textbf{Theorem 2.2.A: ``Bisognano-Wichmann Theorem''} (Bisognano and Wichmann (1975))\\[2pt]
Let $J$ = \textit{modular conjugation}, $\Delta$ = \textit{modular operator} for $\mathsf{N}(W_R)$, $\Omega$.\\
Then the following statements hold. 
 \begin{itemize}
  \item[(1)] $\Delta^{it} = U(0,\Lambda_{2 \pi t})$ \quad $(t \in \mathbb{R})$
  \item[(2)] $ ( \ \Longleftrightarrow (1)\,)$ \ \ The restriction of the vacuum state $\omega_\Omega(\,.\,) = \langle \Omega,\,.\,\Omega \rangle$
  to the vN algebra $\mathsf{N}(W_R)$ is a KMS state at inverse temperature $\beta = 2\pi$ for the automorphic action 
  $$ \gamma_t(A) = U(0,\Lambda_t) A U(0,\Lambda_t)^* \quad \ \ (t \in \mathbb{R}, \ \, A \in \mathsf{N}(W_R) )$$
  of the Lorentz boost group which leaves the wedge region invariant
  \item[(3)] $J \phi(f) J = \phi (\overline{f} \circ \boldsymbol{\iota}) \ \ (\mathrm{supp}(f) \subset W_R)$ 
\ $\Rightarrow \ \ 
  J\mathsf{N}(O) J = \mathsf{N}(\boldsymbol{\iota}(O)) \quad \ \ (O \subset W_R)$
  \\[4pt]  ${}$ \quad \quad \quad $\Rightarrow$ \ \ $\mathsf{N}(W_L) = \mathsf{N}(W_R)'$  \ \ \textbf{wedge duality}
  \item[(4)] ( $\Leftarrow$ (3)) $JU(0,R) = \Theta$ (PCT operator), $R$ = rotation by $\pi$ in $x^2$-$x^3$ plane
 \end{itemize}
 
\subsection{Borchers' Theorem}

The Bisognano-Wichmann Theorem has been generalized in an abstract setting by H.-J.\ Borchers (1992), with simplified proof later given by M. Florig (1998).
\\[6pt]
It takes as starting point the concept of a \textbf{Borchers triple} (named later in this way by D. Buchholz) -- it consists of 
$(\mathsf{N},U,\Omega)$, with the following properties:
\begin{itemize}
 \item[(i)]  $\mathsf{N} \subset \mathsf{B}(\mathcal{H})$ a vN algebra on some Hilbert space $\mathcal{H}$
 \item[(ii)] $U_a$ $(a \in \mathbb{R}^d)$ is a continuous unitary representation on $\mathcal{H}$ fulfilling the \textit{spectrum condition},
 and 
 $$ U_a \mathsf{N} U_a^* \subset \mathsf{N} \ \ \text{for all} \ \ a \in W_R^{(d)} $$
 $W_R^{(d)}$ = $d$-dimensional analogue of $\overline{W_R}$
 \item[(iii)] $\Omega \in \mathcal{H}$ is a \textit{standard vector} for $\mathsf{N}$, with $U_a\Omega = \Omega$
\end{itemize}
\textbf{Theorem 2.3.A: ``Borchers' Theorem''} (Borchers (1992), Florig (1998))\\[2pt]
Let $(\mathsf{N},U,\Omega)$ be a Borchers triple, and $J,\,\Delta$ the modular objects of $\mathsf{N},\Omega$.
\\[2pt]
$$ \text{Then}: \ \quad \ \
 J U_a J = U_{\boldsymbol{\iota}(a)} \quad \text{and} \quad \Delta^{it} U_a \Delta^{-it} = U_{\Lambda_{2 \pi t} a}
 \quad (a \in W_R^{(d)}, \ t \in \mathbb{R}) $$
 ${}$ \\
\textbf{Remarks}
\begin{itemize}
 \item Bisognano-Wichmann Thm \quad $\Longrightarrow$ \quad Borchers Thm \\[2pt]
 The converse does not hold (Yngvason (1994))
 \item KMS states generalize the concept of thermal equilibrium states: \\[2pt]
 If $\mathcal{H}$ is a Hilbert space, $H$ a selfadjoint operator on $\mathcal{D} \subset \mathcal{H}$, \\
 $Z_\beta = \mathrm{Tr}(e^{-\beta H}) < \infty$ $(\beta > 0)$, then 
 $$ \omega^{(\beta)}(A) = \frac{1}{Z_\beta} \mathrm{Tr}(e^{- \beta H} A)  \quad \ \ (A \in \mathsf{B}(\mathcal{H})) $$
 is a thermal equilibrium state for a quantum system with Hamiltonian $H$. 
 \\[2pt]
 $\omega^{(\beta)}$ is a KMS state on $\mathsf{B}(\mathcal{H})$ for the automorphism group $\sigma_t(A) = U_t A U^*_t$ $(A \in \mathsf{B}(\mathcal{H}), \ t \in 
 \mathbb{R})$ with $U_t = e^{itH}$.
 \\[2pt]
 \textit{Note}: The interpretation of a KMS state as thermal equilibrium state requires an inertial frame wherein the 
 ``medium'' is globally at rest, such that $t$ corresponds to the time coordinate. The time parameter $t$ of the Lorentz boosts
 $\Lambda_t$ is not of that form. ($\longrightarrow$ discussions about the ``Unruh effect'')
\end{itemize}
\begin{itemize}
 \item The Bisognano-Wichmann Theorem generalizes to Wightman quantum fields of general spin type (Bisognano and Wichmann (1976)).
 \item For a conformally covariant quantum field theory in Minkowski spacetime, the Tomita-Takesaki modular objects for 
 $\mathsf{N}(O)$, $\Omega$ have geometric significance also if $O$ is a \textbf{double cone}. In this case, they implement
 conformal transformations related to the double cone (Hislop and Longo (1981), Hislop (1988)).
 \item The latter generalizes to de Sitter spacetime (Fr\"ob (2023)).
 \item The question if the Tomita-Takesaki modular objects of $\mathsf{N}(O)$, $\Omega$ have geometric significance
 for non-conformally covariant quantum fields is a longstanding question.
\end{itemize}

\subsection{Araki-Uhlmann Relative Entropy}

The Tomita-Takesaki modular objects, particularly for ``wedge region'' vN algebras, are instrumental for the concept 
 of the \textbf{Araki-Uhlmann relative entropy} in quantum field theory.
 Let $\mathsf{N} \subset \mathsf{B}(\mathcal{H})$ be a vN algebra, $\Omega \in \mathcal{H}$ a standard vector for $\mathsf{N}$,
and let $\Delta$ be the modular operator of $\mathsf{N}$ and $\Omega$.
\\[6pt]
If $U \in \mathsf{N}$ is unitary, then $\Psi = U\Omega$ is again a standard vector for $\mathsf{N}$
\\[6pt]
\textbf{Exercise} \ \ Prove this.
\\[6pt]
The state vectors induce states on $\mathsf{N}$:
$$ \omega_{\Omega}(A) = \langle \Omega,A\Omega \rangle\,, \quad \ \ \ \omega_\Psi(A) = \langle \Psi,A\Psi \rangle \quad \quad (A \in \mathsf{N})$$
Then the \textbf{Araki-Uhlmann relative entropy} is defined as (Araki (1976), Uhlmann (1977))
$$ S(\omega_\Psi | \omega_\Omega) = - \langle U^*\Omega, \log(\Delta) U^*\Omega\rangle $$
This is a special case of a more general definition. It has the advantage that it can be explicitly calculated for certain states of 
interest for CCR- or CAR-algebras. We will use this in Sec.\ 4. 
\\[4pt]
Some authors have suggested that relative entropy is more general than von Neumann/Shannon entropy, and 
that other quantities in thermodynamics or (quantum) information theory can be derived from it.
\newpage \noindent
{\small
\textbf{References}
\\[6pt]
Baumg\"artel, H., Wollenberg, M., \textit{Causal Nets of Operator Algebras: Mathematical Aspects of Algebraic Quantum Field Theory},
Akademie-Verlag, 1992
\\[4pt]
Bratteli, O., Robinson, D.W., \textit{Operator Algebras and Quantum Statistical Mechanics}, Vols.\ 1 and 2, Springer-Verlag, 1987/1997
\\[4pt]
Haag, R., \textit{Local Quantum Physics}, Springer-Verlag, 1996
\\[4pt]
Takesaki, M., \textit{Tomita's Theory of Modular Hilbert Algebras and its Applications}, Lecture Notes Math., Vol.\ 128, Springer-Verlag, 1970
\\[4pt]
Araki, H. (1976), ``Relative entropy of states of von Neumann algebras'', Publ.\ Res.\ Inst.\
Math.\ Sci.\ \textbf{11}, 809–833.
\\[4pt]
Bisognano, J.J., Wichmann, E.H. (1975), ``On the Duality Condition for a Hermitian Scalar Field'',     J.\ Math.\ Phys.\  \textbf{16}, 985-1007
\\[4pt]
Bisognano, J.J., Wichmann, E.H. (1976), ``On the Duality Condition for Quantum Fields'',     J.\ Math.\ Phys.\  \textbf{17}, 303-321
\\[4pt]
Borchers, H-J. (1992), ``The CPT-theorem in two-dimensional theories of local observables'', Commun.\ Math.\ Phys.\ \textbf{143}, 315-332
\\[4pt]
Borchers, H.-J. (2000), ``On revolutionizing quantum field theory with Tomita's modular theory'', 
    J.\ Math.\ Phys.\ \textbf{41}, 3604-3673
\\[4pt]
Florig, M. (1998), ``On Borchers' Theorem'', Lett.\ Math.\ Phys.\ \textbf{46}, 289-293 \\[4pt]
Uhlmann, A. (1977), ``Relative entropy and the Wigner-Yanase-Dyson-Lieb concavity in an
interpolation theory'', Commun.\ Math.\ Phys.\ \textbf{54}, 21–32
\\[4pt]
Yngvason, J. (1994), ``A note on essential duality'',  Lett.\ Math.\ Phys.\ \textbf{31}, 127-141
\\[10pt]
\textbf{Further Reading}
\\[6pt]
Ohya, M., Petz, D., \textit{Quantum Entropy and its Use}, Springer-Verlag, 1993
\\[4pt]
Buchholz, D., Dreyer, O., Florig, M., Summers, S.J. (2000), 
``Geometric modular action and space-time symmetry groups'',
    Rev.\ Math.\ Phys.\ \textbf{12}, 475-560
\\[4pt]
Buchholz, D., Verch, R.  (2016), ``Unruh versus Tolman: On the heat of acceleration'', Gen.\ Rel.\ Grav.\ \textbf{48}, 3, 32
\\[4pt]
Casini, H., Grillo, S., and Pontello, D. (2019), ``Relative entropy for coherent states from
Araki formula'', Phys.\ Rev.\ \textbf{D 99}, 125020 
\\[4pt]
Dowling, N., Floerchinger, S., Haas, T. (2020), ``Second law of thermodynamics for relativistic fluids formulated with relative entropy'',
Phys.\ Rev.\ \textbf{D 102}, 105002 
\\[4pt]
Fr\"ob, M.B. (2023), ``Modular Hamiltonian for de Sitter diamonds'', JHEP \textbf{12}, 074
\\[4pt]
Fr\"ob, M.B., Sangaletti, L. (2025), ``Petz–R\'enyi relative entropy in QFT from modular theory'', Lett.\ Math.\ Phys.\  \textbf{115}, 2, 30
\\[4pt]
Hislop, P.D., Longo, R. (1982), ``Modular structure of the local algebras associated with the free massless scalar field theory'',
    Commun.\ Math.\  Phys.\ \textbf{84}, 71
    \\[4pt]
K\"ahler, R., Wiesbrock, H.W. (2001), ``Modular theory and the reconstruction of four-dimensional quantum field theories'',
    J.\ Math.\ Phys.\ \textbf{42}, 74-86 \\[4pt]
Lechner, G., Sch\"utzenhofer, C. (2014), ``Towards an operator-algebraic construction of integrable global gauge theories'',
    Annales Henri Poincar\'e  \textbf{15}, 645-678
\\[4pt]
Passegger, A.G., Verch, R. (2025), ``Disjointness of inertial KMS states and the role of Lorentz symmetry in thermalization'',
    Rev.\ Math.\ Phys.\ \textbf{37}, 2430009
}
\newpage \noindent
\section{Local Covariant Quantum Field Theory}

\subsection{Local covariant quantum field theory as a functor}

In this section, we largely draw on material from Brunetti, Fredenhagen and Verch (2003) and from Fewster and Verch (2016),
and give an introduction into local covariant quantum field theory. One of the motivations comes from attempts to generalize 
Einstein's field equations of gravity to a semiclassical form, where gravity is still described in terms of classical spacetime 
geometry, but coupled to matter described in terms of quantized fields. (A more extended discussion of the line of thought which 
we will now present appears in Wald (1994).) The \textbf{semiclassical Einstein equations} then 
take, at least at the level of an ansatz, the form 
$$ R_{ab}^{[g]}(p) - \frac{1}{2}R^{[g]}g_{ab}(p) = 8\pi \langle T_{ab}^{[\phi,g]}(p)\rangle_\omega  \quad \ \ (p \in M)$$
On the left hand side we have the Einstein tensor of a spacetime $(M,g)$ with spacetime manifold $M$ and metric $g$ (more fully
written as $g_{ab}$ in abstract index notation), and on the right hand side, the \textbf{expectation value of the stress-energy tensor of
a quantum field $\phi$ on the spacetime $(M,g)$ in a state $\omega$}. There are many issues one may want to discuss about the 
possible range of validity of such an equation (see, e.g.\, Meda, Pinamonti and Siemmssen (2021)). However, the point to be addressed here 
is that, if one wishes to impose such an equation at all, then one needs to make sense of the right hand side to begin with. 
In particular, the left hand side behaves covariantly under diffeomorphisms, in the following sense: If $\psi : (M,g) \to (\tilde{M},\tilde{g})$ is an isometry between two spacetimes, then it holds that $\psi_*g = \tilde{g}$, where $\psi_* f = f\circ \psi^{-1}$ is 
the push-forward map turning functions on $M$ into functions on $\tilde{M}$, and we denote the tensorial push-forward map again by 
$\psi_*$. Covariance now means $\psi_*( R_{ab}^{[g]} - \frac{1}{2}R^{[g]}g_{ab}) = R_{ab}^{[\tilde{g}]} -\frac{1}{2}R^{[\tilde{g}]}\tilde{g}_{ab}$. The right hand side of the semiclassical Einstein equation above must, for consistency, obey the same covariance property, 
meaning that $ \psi_*(\langle T_{ab}^{[\phi,g]}\rangle_\omega) = \langle T_{ab}^{[\tilde{\phi},\tilde{g}]}\rangle_{\tilde{\omega}}$
where the quantum field $\tilde{\phi}$ on $(\tilde{M},\tilde{g})$ is the ``counterpart'' of the quantum field $\phi$ on 
$(M,g)$, and $\tilde{\omega}$ is a suitable ``counterpart'' of $\omega$. A way of ensuring the covariance property at the algebra level is by assuming the there is a unital 
$*$-algebra isomorphism $\alpha_\psi$ mapping the algebra of observables of $\phi$ onto the algebra of observables of $\tilde{\phi}$,
so that $\alpha_\psi(T_{ab}^{[\phi,g]}) = \psi_*(T_{ab}^{[\phi,g]})$, and $\tilde{\omega} = \omega \circ \alpha_\psi^{-1}$.
\\[6pt]
This idea can now be formalized and extended, leading to the concept of a local covariant quantum field theory. We start with a basic 
definition. 
\\[6pt]
Let $(M,g)$ and $(\tilde{M},\tilde{g})$ be two 4-dimensional, globally hyperbolic spacetimes, both carrying an orientation and time-orientation. An injective $C^\infty$ map $\psi : M \to \tilde{M}$ is called a \textbf{hyperbolic embedding} if 
\begin{itemize}
 \item $\psi$ is an \textbf{isometry}, i.e.\ $\psi_*g = \tilde{g}$
 \item $\psi$ preserves orientation and time-orientation
 \item If $p$ and $q$ are causally related points ($q \in J^+(p)$ or $q \in J^-(p)$) in $(M,g)$, then every causal curve 
 connecting $\psi(q)$ and $\psi(p)$ in $(\tilde{M},\tilde{g})$ lies in $\psi(M)$.
\end{itemize}
${}$ \vspace*{-1.6cm}
\begin{center}
\includegraphics[width=10.5cm]{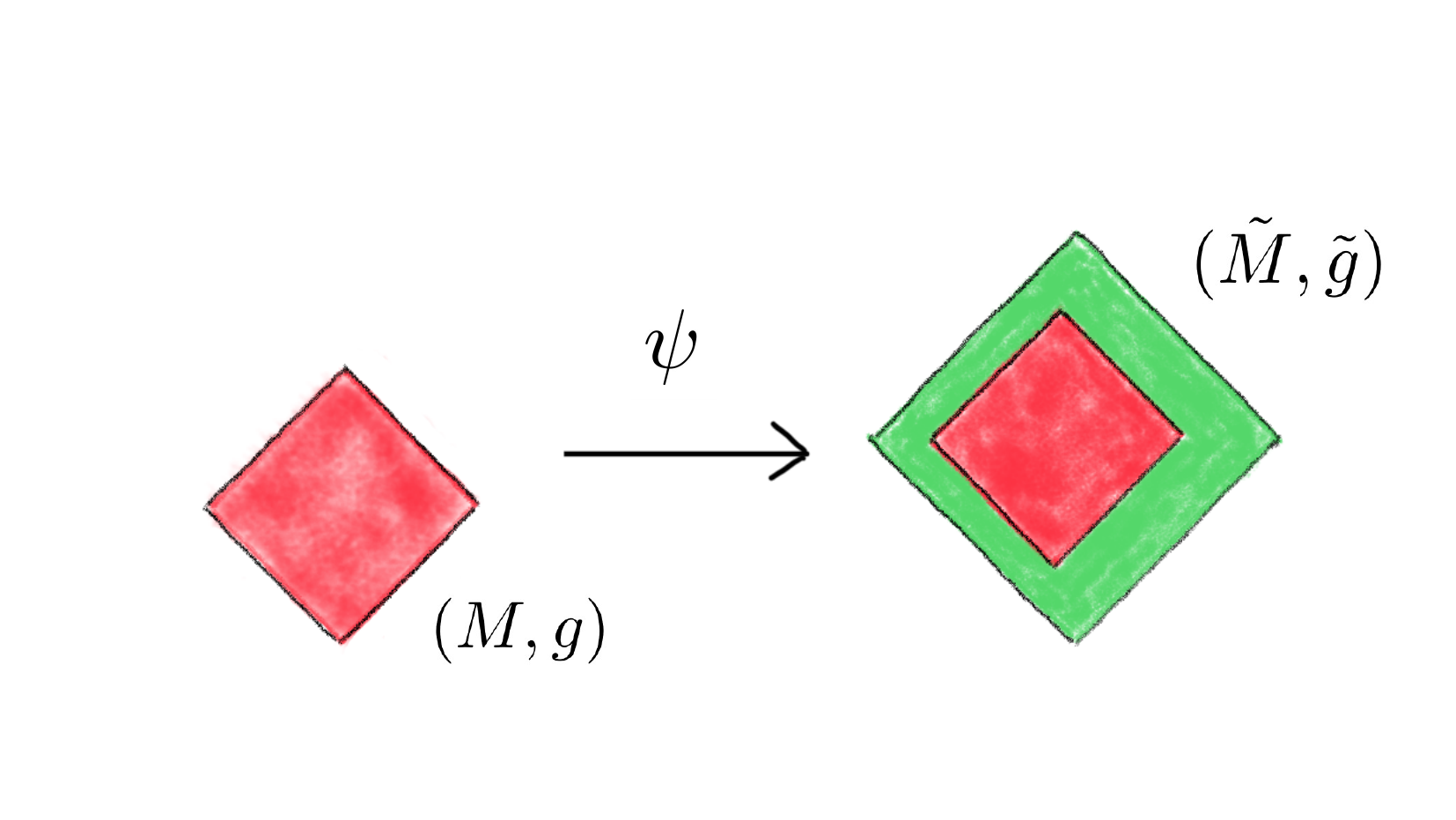}
\end{center}
{\small \textbf{Figure 3.} A hyperbolic embedding $\psi$ of a globally hyperbolic spacetime $(M,g)$ (usually depicted as a double cone) into a larger one $(\tilde{M},\tilde{g})$}
\\ \\
\begin{center}
\includegraphics[width=10.5cm]{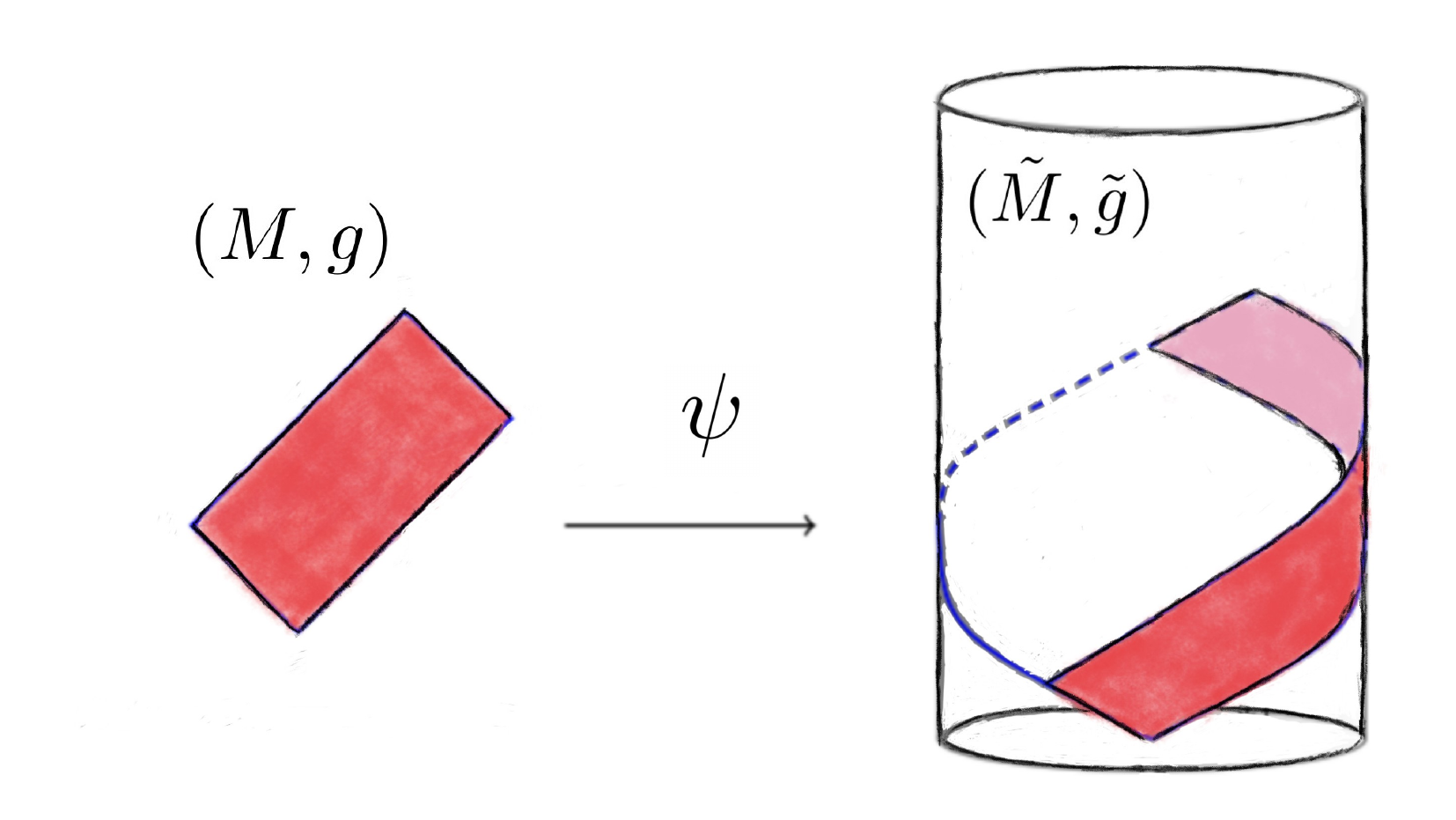}
\end{center}
{\small
\textbf{Figure 4.} An isometric embedding $\psi$ of a globally hyperbolic spacetime $(M,g)$ into a globally hyperbolic $(\tilde{M},\tilde{g})$, the latter a ``timelike cylider spacetime'' with compact Cauchy-surfaces. The embedded image $\psi(M)$ ``wraps around the cylinder'', which allows for a causal curve in $(\tilde{M},\tilde{g})$, depicted by the blue line (drawn broken as it winds around the  ``reverse side'' of the cylinder), that is not in $\psi(M)$ but connects points in $\psi(M)$. The figure is adapted from a similar figure in 
Hollands and Wald (2015).}
\\ \\
The idea behind the concept of a local covariant quantum field theory is to generalize the isotony property of a Haag-Kastler net 
of observables on a fixed spacetime to an assignment of spacetimes to algebras, where the inclusion of spacetime regions is 
generalized to hyperbolic embeddings of spacetimes, and inclusion of algebras to monomorphic embeddings. Here, an injective unital $*$-algebra morphism $\alpha : \mathbfcal{A}_1 \to \mathbfcal{A}_2$ of two unital $*$-algebras is called a \textbf{momomorphic embedding}. 
\\ \\
\textbf{Definition}
\\[2pt]
A \textbf{local covariant quantum field theory} $\mathscr{A} = (\mathcal{A},\alpha)$ consists of 
\begin{itemize}
 \item An assignment $(M,g) \to \mathcal{A}(M,g)$ of a unital $*$-algebra $\mathcal{A}(M,g)$ to every globally hyperbolic 
 spacetime $(M,g)$ 
 \item An assignment $\psi \to \alpha_\psi$ of a monomorphic embedding $\alpha_{\psi} : \mathcal{A}(M,g) \to \mathcal{A}(\tilde{M},\tilde{g})$ to every hyperbolic embedding $\psi : (M,g) \to (\tilde{M},\tilde{g})$, such that the following covariance property holds:
 $$ \alpha_{\psi_2} \circ \alpha_{\psi_1} = \alpha_{\psi_2 \circ \psi_1} $$
 together with $\alpha_{\iota_{M,M}}(\Att) = \Att$ $(\Att \in \mathcal{A}(M,g))$, where $\iota_{M,M}: p \mapsto p$ $(p \in M)$ is the identity map on $M$.
\end{itemize}
At this point, it is straightforward to notice that these assignments have the structure of a \textbf{functor} between the 
following two categories:
\\[6pt]
$\mathbb{GH}${\scriptsize$\mathbb{YP}$} is the category whose objects are 4-dimensional globally hyperbolic spacetimes $(M,g)$ with orientation and time-orientation, and whose morphisms (or arrows) are hyperbolic embeddings $(M,g) \overset{\psi}{\longrightarrow} (\tilde{M},\tilde{g})$
\\[6pt]
$\mathbb{A}${\scriptsize$\mathbb{LG}$} is the category whose objects are unital $*$-algebras $\mathcal{A}$, and whose morphisms (or arrows) are monomorphic embeddings $\mathcal{A}_1 \overset{\alpha}{\longrightarrow} \mathcal{A}_2 $
\\[6pt]
The functorial property can then be conveniently expressed in a diagrammatic form,
$$
\begin{CD}
{(M,g)} @>\psi>> {(\tilde{M},\tilde{g})}   \\
@V{\mathscr{A}}VV     @VV{\mathscr{A}}V   \\
{\mathscr{A}}(M,g)@>{{\mathscr{A}}(\psi)}>> {\mathscr{A}}(\tilde{M},\tilde{g})
\end{CD}
$$
where we used the usual convention in functor notation to denote the assignments of objects and of morphisms with the same functor symbol, i.e.\
$\mathscr{A}(M,g) = \mathcal{A}(M,g)$ and $\mathscr{A}(\psi) = \alpha_\psi$. The diagram can be enlarged in an obvious way to express the 
covariance property $\mathscr{A}(\psi_2) \circ \mathscr{A}(\psi_1) = \mathscr{A}(\psi_2 \circ \psi_1)$ --- we leave this as an \textbf{exercise}.
${}$\\[6pt]
There are further properties that are usually imposed on a local covariant quantum field theory:
\\[8pt]
\textbf{Causality}: If there are any two (or more generally, finitely many) hyperbolic embeddings $\psi_1 : (M_1,g_1) \to (\tilde{M},\tilde{g})$ and $\psi_2: (M_2,g_2) \to (\tilde{M},\tilde{g})$ such that $\psi_1(M_1)$ and $\psi_2(M_2)$ are \textbf{causally separated} in $(\tilde{M},\tilde{g})$ (i.e.\ there is no causal curve in $(\tilde{M},\tilde{g})$ connecting $\psi_1(M_1)$ and $\psi_2(M_2)$), then 
the sub-$*$-algebras $\mathscr{A}(\psi_1)(\mathscr{A}(M_1,g_1))$ and $\mathscr{A}(\psi_1)(\mathscr{A}(M_2,g_2))$ of $\mathscr{A}(\tilde{M},\tilde{g})$ commute. 
${}$\\[6pt]
\begin{center}
\includegraphics[width=12cm]{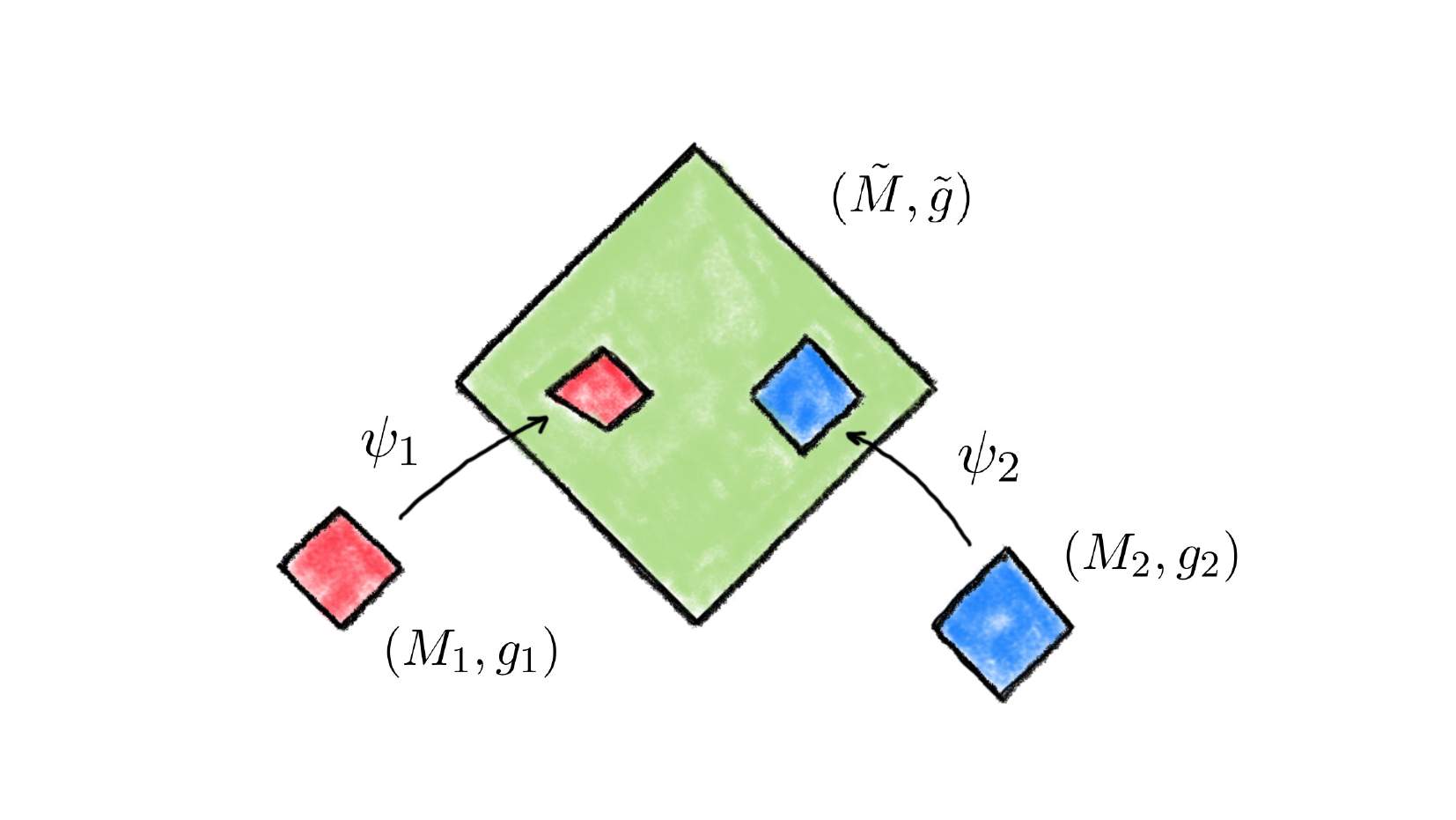}
\end{center}
{\small
\textbf{Figure 5.} Two hyperbolic embeddings $\psi_1 : (M_1,g_1) \to (\tilde{M},\tilde{g})$ and $\psi_2: (M_2,g_2) \to (\tilde{M},\tilde{g})$ such that $\psi_1(M_1)$ and $\psi_2(M_2)$ are {causally separated} in $(\tilde{M},\tilde{g})$.}
\\ \\
\textbf{Time-slice property}: If for any hyperbolic embedding $\psi: (M,g) \to (\tilde{M},\tilde{g)}$ the embedded image $\psi(M)$ contains a Cauchy-surface for $(\tilde{M},\tilde{g})$, then $\mathscr{A}(\psi)(\mathscr{A}(M,g)) = \mathscr{A}(\tilde{M},\tilde{g})$.
\\[6pt]
\begin{center}
\includegraphics[width=10.5cm]{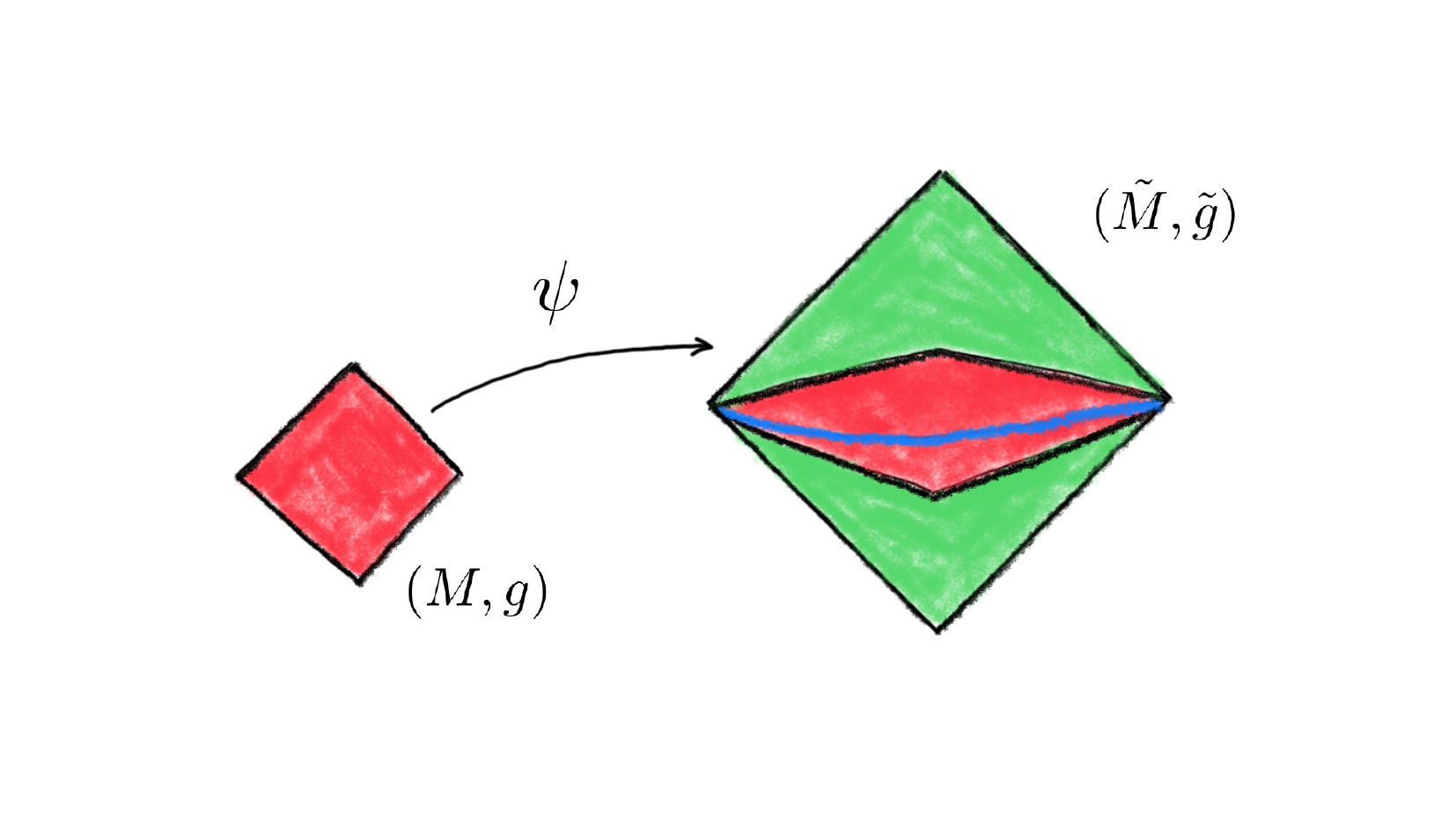}
\end{center}
{\small
\textbf{Figure 6.} A hyperbolic embedding $\psi : (M,g) \to (\tilde{M},\tilde{g})$ so that $\psi(M)$ contains a Cauchy-surface of $(\tilde{M},\tilde{g})$, indicated by the bold blue line.} 
\\ \\
\textbf{Remarks}
\begin{itemize}
 \item Here, we have taken $\mathbb{A}${\scriptsize$\mathbb{LG}$} as the target category of the functor $\mathscr{A}$, comprised by unital $*$-algebras and monomorphisms between them. In some contexts, it is more useful to consider instead the category 
 $C^*$-$\mathbb{A}${\scriptsize$\mathbb{LG}$} which has unital-$C^*$-algebras as objects, and unital $C^*$-monomorphisms as 
 arrows.
 \item Simple examples for local covariant quantum field theories are the quantized scalar Klein-Gordon field, the Proca field and 
 the Dirac field (for the latter, also spin-structures on globally hyperbolic spacetimes have to be taken into account, and hyperbolic
 embeddings are required to preserve spin struture, cf.\ Verch (2001), Sanders (2010)). 
 \item Among others, the local covariant approach 
 allows it to derive a spin-statistics theorem for quantum field theory in curved spacetime (Verch (2020)). 
 \item Interacting quantum field theories can be constructed in the sense of perturbation theory and formulated in a unital $*$-algebraic setting. These constructions also provide examples of local covariant quantum field theories, e.g.\ scalar fields with $\phi^p$ self-interaction, and Yang-Mills theories (Hollands (2008)). See Rejzner (2016) and Hollands and Wald (2015) and literature cited there for further discussion.
 \item Quantized free electrodynamics can also be constructed as a local covariant quantum field theory, but subtleties arise since
 the quantized electrodynamic potential is sensitive to the global topology of the spacetime. For a discussion of these issues,
 see e.g.\ Sanders, Dappiaggi and Hack (2024)
 \item A novel $C^*$-algebraic construction of interacting quantum field theories has recently been proposed in Buchholz and Fredenhagen (2020) (in the entries of Further Reading). This construction can also be employed for quantum field theory in curved spacetimes.
\end{itemize}
\subsection{Haag-Kastler nets on a fixed spacetime from a local covariant QFT}

Let $\mathscr{A}$ be local covariant quantum field theory.
\\[6pt]
Consider an object $(M,g)$ in $\mathbb{GH}${\scriptsize$\mathbb{YP}$}. 
\\[2pt]
Denote by $\mathrm{Iso}^\uparrow_+(M,g)$ the group of the isometries of $(M,g)$ that preserve time-orientation and 
orientation.
\\[2pt]
Suppose that $O$ is an open subset of $M$ and that (i) $(O,g|_O)$ is an object in $\mathbb{GH}${\scriptsize$\mathbb{YP}$}, where 
$g|_O$ denotes the restriction of the metric $g$ to $O$, and (ii) the identical embedding map $\iota_{M,O} : O \to M$, $\iota_{M,O}(p) = p$ $(p \in O)$ is an arrow in $\mathbb{GH}${\scriptsize$\mathbb{YP}$}. We define unital sub-$*$-algebra of $\mathscr{A}(M,g)$, 
$$ \mathcal{A}(M,O) = \mathscr{A}(\iota_{M,O})(\mathscr{A}(O,g|_O)) $$
Then the following result obtains.
\\[6pt]
\textbf{Theorem 3.2.A} (Brunetti, Fredenhagen and Verch (2003))
\begin{itemize}
 \item[(1)] $O_1 \subset O_2 \Longrightarrow \mathcal{A}(M,O_1) \subset \mathcal{A}(M,O_2)$
 \item[(2)] There is a representation $\alpha_\kappa$ $(\kappa \in \mathrm{Iso}^\uparrow_+(M,g))$ of 
 $\mathrm{Iso}^\uparrow_+(M,g)$ by automorphisms of $\mathcal{A}(M,g)$ with the covariance property 
 $$ \alpha_{\kappa}(\mathcal{A}(M,O)) = \mathcal{A}(M,\kappa(O))  $$

 \item[(3)] If $O_1$ and $O_2$ are causally separated $(O_1 \cap J^\pm(O_2) = \emptyset)$ in $(M,g)$, then 
 the algebras $\mathcal{A}(M,O_1)$ and $\mathcal{A}(M,O_2)$ commute. 
 \item[(4)] The time-slice axiom holds: If $O \subset \tilde{O}$ are open subsets of $M$ and 
 $(O,g|_O)$ and $(\tilde{O},g|_{\tilde{O}})$ are objects in $\mathbb{GH}${\scriptsize$\mathbb{YP}$} such that
 $O$ is an open neighbourhood, in $\tilde{O}$, of a Cauchy-surface 
 for $(\tilde{O},g|_{\tilde{O}})$, then
 $$\mathcal{A}(M,O) = \mathcal{A}(M,\tilde{O}) $$
\end{itemize}
We will next indicate the arguments for the properties (1) and (2). 
Properties (3) and (4) are clearly derived from causality and time-slice property of $\mathscr{A}$. This makes a good 
\textbf{exercise}. 
\\[6pt]
\textit{Proof}. (1) The embeddings $\iota_{M,O} : (O,g|_O) \to (M,g)$, $\iota_{M,\tilde{O}} : (\tilde{O},g|_{\tilde{O}}) \to (M,g)$ and 
$\iota_{\tilde{O},O} :(O,g|_O) \to (\tilde{O},g|_{\tilde{O}})$ given by the identity maps $p \mapsto p$ with the corresponding domains and ranges fulfill the composition property $\iota_{M,O} = \iota_{M,\tilde{O}}\circ \iota_{\tilde{O},O}$. Hence, we obtain
\begin{align*}
 \mathcal{A}(M,O) = \mathscr{A}(\iota_{M,O})(\mathscr{A}(O,g|_O)) & = \mathscr{A}(\iota_{M,\tilde{O}}) \circ \mathscr{A}(\iota_{\tilde{O},O})(\mathscr{A}(O,g|_O)) \\
 & \subset \mathscr{A}(\iota_{M,\tilde{O}})(\mathscr{A}(\tilde{O},g|_{\tilde{O}})) = \mathcal{A}(M,\tilde{O})
\end{align*}
since $\mathscr{A}(\iota_{\tilde{O},O})(\mathscr{A}(O,g|_O)) \subset \mathscr{A}(\tilde{O},g|_{\tilde{O}})$. 
\\[2pt]
(2) Let $\kappa \in \mathrm{Iso}_+^\uparrow(M,g)$. Then $\kappa : (M,g) \to (M,g)$ is an arrow in $\mathbb{GH}${\scriptsize$\mathbb{YP}$}, and likewise so is $\kappa^{-1} : (M,g) \to (M,g)$, and the identity map $\iota_{M,M}$. 
Setting $\alpha_\kappa = \mathscr{A}(\kappa)$, it holds that $\alpha_\kappa \circ \alpha_{\kappa'} = \alpha_{\kappa \circ \kappa'}$ and $\alpha_{\kappa} \circ \alpha_{\kappa^{-1}} = \alpha_{\iota_{M,M}}$ by the properties of the functor 
$\mathscr{A}$. Moreover, by the time-slice property, the $\alpha_\kappa$ are bijective, and we have $\mathscr{A}(\iota_{M,M}) = \mathrm{id}_{\mathscr{A}(M,g)}$, the identical map on $\mathscr{A}(M,g)$, as a requirement on $\mathscr{A}$. Taking this together, $\kappa \mapsto \alpha_\kappa$ is a representation of $\mathrm{Iso}_+^\uparrow(M,g)$ by automorphisms of $\mathscr{A}(M,g)$.
\\[2pt]
Assuming that $O \subset M$ is open and $ \iota_{M,O} : (O,g|_O) \to (M,g)$ is an arrow in $\mathbb{GH}${\scriptsize$\mathbb{YP}$}, one has 
\begin{align*}
 \alpha_\kappa(\mathcal{A}(M,O)) & = \mathscr{A}(\kappa) \circ \mathscr{A}(\iota_{M,O})(\mathscr{A}(O,g|_O)) \quad \ \ \ \text{and} \\
 \mathcal{A}(M,\kappa(O)) & = \mathscr{A}(\iota_{M,\kappa(O)})(\mathscr{A}(\kappa(O),g|_{\kappa(O)}) \\
                        & = \mathscr{A}(\iota_{M,\kappa(O)}) \circ \mathscr{A}(\kappa_O)(\mathscr{A}(O,g|_{O}))
\end{align*}
where $\kappa_O : (O,g|_O) \to (\kappa(O),g|_{\kappa(O)})$, $p \mapsto \kappa(p)$ is the restriction of $\kappa$ in 
domain and range, inducing an arrow in $\mathbb{GH}${\scriptsize$\mathbb{YP}$}. However, we have 
$\mathscr{A}(\kappa) \circ \mathscr{A}(\iota_{M,O}) = \mathscr{A}(\iota_{M,\kappa(O)}) \circ \mathscr{A}(\kappa_O)$ by the 
functorial properties of $\mathscr{A}$, showing that $\alpha_\kappa(\mathcal{A}(M,O)) = \mathcal{A}(M,\kappa(O))$.
${}$ \hfill $\Box$
\\[6pt]
The result shows that any local covariant QFT induces an algebraic QFT on any fixed globally hyperbolic spacetime, in the form of a Haag-Kastler net of unital $*$-algebras (or unital $C^*$-algebras, as the proof remains unchanged in this case). 

\subsection{Relative Cauchy evolution: Dynamics from embeddings and time-slice property}

We consider the case that we are given two globally hyperbolic spacetimes: $(M,g)$ and $(M,g[h])$, where 
$g[h]$ differs from $g$ by a ``perturbation'' $h$ which is confined to a compact set. That means, $g[h] = g + h$ where 
$h$ is a compactly supported smooth 2-cotensor-field, such that $g[h]$ is again a globally hyperbolic Lorentzian metric on 
$M$. Then there are joint Cauchy-surfaces $\Sigma_\pm$ for $(M,g)$ and $(M,g[h])$ so that $\mathrm{supp}(h) \subset J^-(\Sigma_+)$ and $\mathrm{supp}(h) \subset J^+(\Sigma_-)$. Thus, $g = g[h]$ on $M^\pm = \mathrm{int}(J^\pm(\Sigma_\pm))$
\footnote{$\mathrm{int}(X)$ is the open interior of a subset $X$ of a topological space}
\\[2pt]
It will now be convenient to change the notation for the globally hyperbolic spacetimes (i.e.\ the objects in 
$\mathbb{GH}${\scriptsize$\mathbb{YP}$}) arising in the situation at hand, as follows:
$$ \Mb = (M,g)\,, \quad \ \ \Mb [h] = (M,g[h])\,, \quad \ \ \Mb^\pm =(M^\pm,g|_{M^\pm}) $$
It is worth noting that $M^\pm$ contain Cauchy-surfaces for $(M,g)$ and $(M,g[h])$. 
\\[2pt]
There are the hyperbolic embeddings $\iota^\pm: \Mb^\pm \to \Mb$ and $j^\pm:\Mb \to \Mb[h]$ given by the 
identical maps $p \mapsto p$; they are arrows in 
$\mathbb{GH}${\scriptsize$\mathbb{YP}$}, and their ranges contain Cauchy-surfaces.
\begin{center}
 \includegraphics[width=10.6cm]{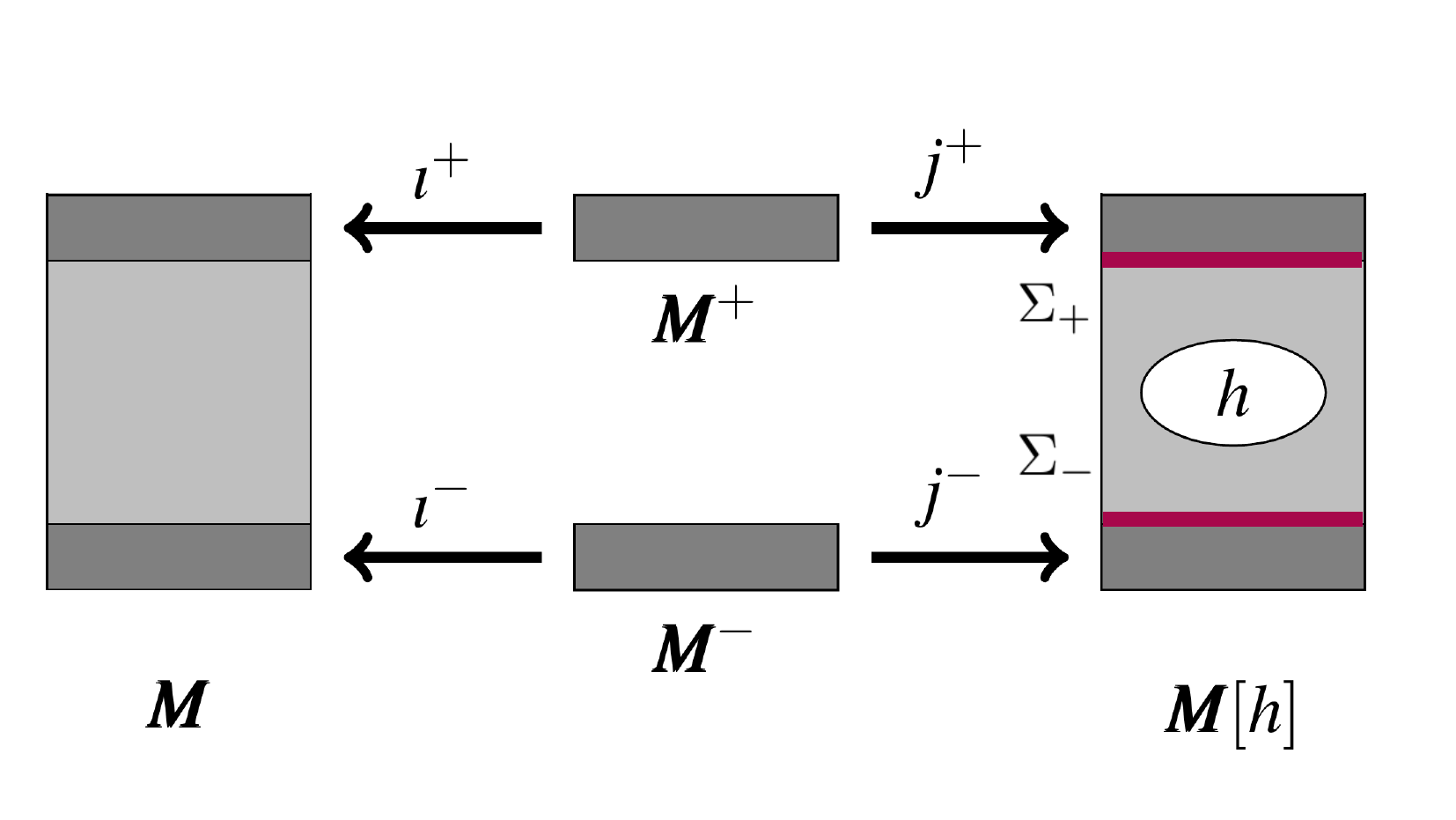}
\end{center}
{\small
\textbf{Figure 7.} Illustration of the spacetimes $\Mb$ and $\Mb[h]$ together with the embeddings of $\Mb^\pm$ described in the text. For $\Mb[h]$, the Cauchy-surfaces $\Sigma_\pm$ are indicated by bold purple lines. Adapted from a 
similar figure in Fewster and Verch (2015).
}
\\ \\
Now let $\mathscr{A}$ be a local covariant QFT. The above defined embeddings then have functorial counterparts under $\mathscr{A}$, depicted as follows:
\begin{center}
 \includegraphics[width=16cm]{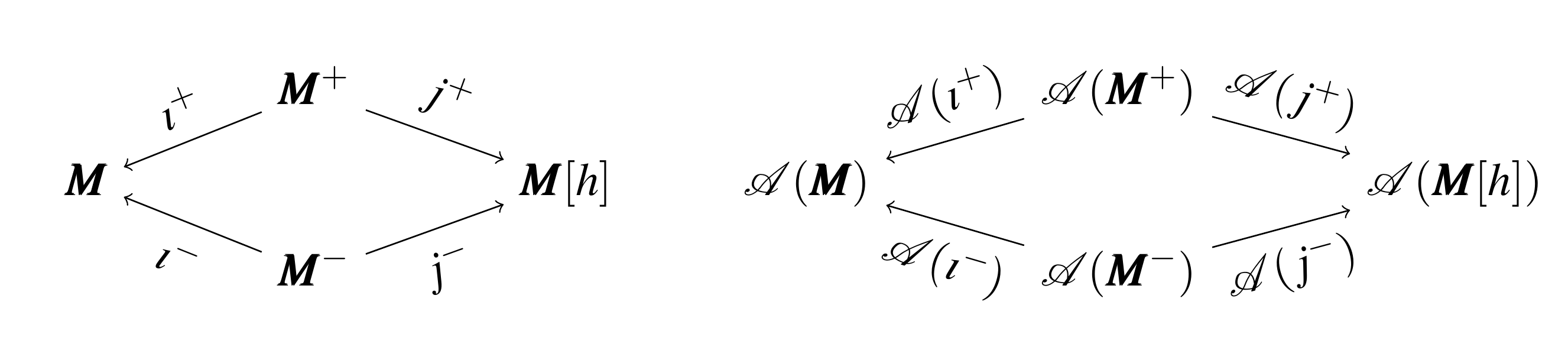}
\end{center}
{\small
\textbf{Figure 8.} Diagrammatic representation of the hyperbolic embeddings $\iota^\pm$ and $j^\pm$ (left) and the 
corresponding arrows in $\mathbb{A}${\scriptsize$\mathbb{LG}$} assigned to the former by the functor $\mathscr{A}$ (right). The figure is taken from Fewster and Verch (2015).
}
\\ \\
By the time-slice property, all the algebra morphisms are bijective, hence one can form 
the \textbf{relative Cauchy evolution}
$$ \mathrm{rce}_{\boldsymbol{M}}[h] = \mathscr{A}(\iota^-) \circ \mathscr{A}(j^-)^{-1} \circ \mathscr{A}(j^+) \circ \mathscr{A}(\iota^+)^{-1} $$
The relative Cauchy evolution (sometimes also called ``scattering morphism'' in other contexts) describes the 
change in dynamics resulting from the metric perturbation $h$, compared to the dynamics without the metric perturbation in 
the spacetime $\Mb = (M,g)$ with the ``unperturbed'' background metric $g$. In a Hilbert space representation of $\mathscr{A}(\Mb)$, it can be viewed as the adjoint action of a unitary $S$-matrix on the representation Hilbert space, and there 
are examples of local covariant QFTs where the scattering of a quantum field by an external spacetime metric perturbation 
is indeed described by an $S$-matrix. Further below, we will again come back to that point. 
\\[6pt]
We collect some properties of the relative Cauchy evolution for a local covariant QFT $\mathscr{A}$.
\\[6pt]
\textbf{Proposition 3.3.A} 
\begin{itemize}
 \item[(A)] If $\psi: \Mb \to \widetilde{\Mb}$ is an arrow in $\mathbb{GH}${\scriptsize$\mathbb{YP}$}, then 
 $$ \mathrm{rce}_{\widetilde{\boldsymbol{M}}}[\psi_*h] \circ \mathscr{A}(\psi) = \mathscr{A}(\psi) \circ \mathrm{rce}_{\boldsymbol{M}}[h] $$
 \item[(B)] Suppose that $\mathscr{A}$ is the local covariant QFT of the quantized Klein-Gordon field with field 
 equation $(\Box + m^2)\varphi = 0$ on every globally hyperbolic spacetime, where the constant $m \ge 0$ is the same 
 for every spacetime. Then there is for every globally hyperbolic spacetime $\Mb$ a dense unital sub-$*$-algebra $\mathcal{P}(\Mb)$ of $\mathscr{A}(\Mb)$ so that 
 $$ \left.\frac{1}{2i}\frac{d}{d\lambda}\omega(\mathrm{rce}_{\boldsymbol{M}}[\lambda h](\Att)) \right|_{\lambda = 0}
 = \omega([T_{\boldsymbol{M}}(h),\Att])\,, \quad \ \ (\Att \in \mathcal{P}(\Mb)) $$
 holds for all Hadamard states $\omega$ of $\mathscr{A}(\Mb)$, where $T_{\boldsymbol{M}}(h)$ is the operator-valued stress-energy tensor of the 
 quantized Klein-Gordon field on $\Mb$, smeared with the metric-pertur\-bation $h$.
\end{itemize}
The notation and statements of (B) are somewhat symbolic and we refer to Brunetti, Fredenhagen and Verch (2003) for 
full details, as well as for the proof. Statement (A) is taken from Fewster and Verch (2012). 

\subsection{States for $\mathscr{A}$}

Let $\mathscr{A}$ be a local covariant QFT. Then a \textbf{state} for $\mathscr{A}$ is, by definition, a family
$\boldsymbol{\omega} = \{\omega_{\boldsymbol{M}}\}$, where $\Mb$ ranges over all objects of $\mathbb{GH}${\scriptsize$\mathbb{YP}$}, such that $\omega_{\boldsymbol{M}}$ is a state on $\mathscr{A}(\Mb)$.
\\[6pt]
This is not the only possible notion of state for $\mathscr{A}$. It leans towards the concept of viewing a co-vectorfield
$\xi$ on a manifold $Y$ as inducing a functional on every tangent space $T_yY$ through 
$\xi_y: v_y \mapsto \xi_y(v_y)$ for every $v_y \in T_yY$ $(y \in Y)$. The analogue is seen by regarding
the manifold point $y$ as playing the role of $\Mb$ in the object class of $\mathbb{GH}${\scriptsize$\mathbb{YP}$}, and
$T_yY$ as playing the role of $\mathscr{A}(\Mb)$.
\\[6pt]
In quantum field theory on a fixed globally hyperbolic spacetime, it is usually assumed that there is an 
$\mathrm{Iso}_+^\uparrow(\Mb)$-invariant state $\omega_0$, and that this state can be regarded as a 
physical state (potentially under imposing further conditions). In fact, this point of view is corroborrated by 
many examples. Recall that typically, for a QFT on a fixed globally hyperbolic spacetime given in terms of 
local algebras $\mathcal{A}(M,O)$ we assume that there is a representation $\alpha_\kappa$ $(\kappa \in \mathrm{Iso}_+^\uparrow(M,g))$ of the isometry group by automorphisms of $\mathcal{A}(M)$, with $\alpha_\kappa(\mathcal{A}(M,O)) = \mathcal{A}(M,\kappa(O))$. As discussed above, a local covariant QFT induces such a structure for every 
globally hyperbolic spacetime $(M,g)$. To say that the state $\omega_0$ is $\mathrm{Iso}^\uparrow_+(\Mb)$-invariant then means that 
$\omega_0 \circ \alpha_\kappa = \omega_0$ holds for all $\kappa \in \mathrm{Iso}^\uparrow_+(\Mb)$. 
\\[6pt]
Now the question arises if the concept of an invariant state can be generalized to the previously given notion
of states for a local covariant QFT $\mathscr{A}$. A generalization which seems immediate is as follows: One defines
$\boldsymbol{\omega} = \{\omega_{\boldsymbol{M}}\}$ to be an \textbf{invariant state} (sometimes also called 
\textbf{natural state}) if 
$$ \omega_{\widetilde{\boldsymbol{M}}} \circ \mathscr{A}(\psi) = \omega_{\boldsymbol{M}} $$
holds for all arrows $\psi: \Mb \to \widetilde{\Mb}$. 
\\[6pt]
As ``natural'' as this concept may appear at a first glance,
it turns out that in great generality, \textit{local covariant QFTs do not 
admit invariant states} in this sense. Already the local covariant QFT of the quantized Klein-Gordon field provides
a simple counterexample, as observed in Brunetti, Fredenhagen and Verch (2003). To sketch the argument,
take $\Mb$ to be Minkowski spacetime. Then assuming that we have an invariant state for the 
local covariant QFT of the quantzed Klein-Gordon field, its state $\omega_{\boldsymbol{M}}$ on $\mathscr{A}(\Mb)$
must obviously be equal to $\omega_0$, the $\mathrm{Iso}^\uparrow_+(\Mb)$-invariant vacuum state on Minkowski spacetime.\footnote{Strictly speaking, the vacuum state fulfills, additionally to invariance under isometries, also 
the spectrum condition, but clearly this is a condition that one would like to include to ensure a physical state.}
On the other hand, every invariant state for a local covariant QFT must also be invariant under any relative Cauchy evolution. In the case of the quantized Klein-Gordon field at hand, this means that 
$$ \omega_0 \circ \mathrm{rce}_{\boldsymbol{M}}[h] = \omega_0 $$
for all compactly supported metric perturbations $h$, with $\Mb$ = Minkowski spacetime. However, this last equation 
fails to hold: It has been shown in Wald (1979) that in the GNS representation of $\omega_0$, $\mathrm{rce}_{\boldsymbol{M}}[h]$ is given by the adjoint action of a unitary $S$-matrix, and the $S$-matrix is different from a multiple of the unit operator. In fact, the action of the $S$-matrix on the GNS vector of $\omega_0$ has been calculated in Wald (1979), showing that the invariance of 
$\omega_0$ under the relative Cauchy evolution cannot hold. In the analogy mentioned above where a state for a local covariant QFT is compared to a co-vectorfield on a manifold, the requirement of invariance is akin to wishing for 
an everywhere constant co-vectorfield while the manifold isn't flat.
\\[6pt]
We mention that there is a more general, model-independent no-go theorem concerning the possibility of invariant states
for a local covariant QFT. For full details on that matter, we refer to Fewster and Verch (2015).

\subsection{State space}

The no-go result for invariant states underlines the difficulty of specifying distinguished
states for local covariant QFTs. This is a known issue in QFT on curved spacetimes, making attempts of 
generalizing particle concepts familiar from QFT on Minkowski spacetime to more general spacetimes difficult.
\\[6pt]
It turns out that it is nevertheless possible to introduce the concept of a state space, meaning a set of states 
that can be viewed as physical states, for local covariant quantum field theories. In fact, it can be described
in a functorial setting, similar to (or rather: dual to) that of local covariant quantum field theory.
However, we will not elaborate on the functorial aspects here and refer to Brunetti, Fredenhagen and Verch (2003) for more 
details in that direction. 
\\[6pt]
A \textbf{state space} for a local covariant QFT $\mathscr{A}$ is an assignment $\Mb \mapsto \mathscr{S}(\Mb)$ that 
assigns to every globally hyperbolic spacetime a subset $\mathscr{S}(\Mb)$ of the set of all states on $\mathscr{A}(\Mb)$. 
We recall that a state on $\mathscr{A}(\Mb)$ is a linear functional $\omega: \mathscr{A}(\Mb) \to \mathbb{C}$ that 
fulfills $\omega(\Att^*\Att) \ge 0$ and $\omega(\mathbf{1}) = 1$; if $\mathscr{A}(\Mb)$ is not a $C^*$-algebra, then 
one might add some continuity condition on $\omega$ in order to qualify as ``state''. 
\\[6pt]
A recurrent theme in quantum field theory in curved spacetime, as alluded to before, is the question how to characterize ``physical states''. The ``state space'' for $\mathscr{A}$ should be determined such that the elements in 
$\mathscr{S}(\Mb)$ can be interpreted as physical states for $\mathscr{A}(\Mb)$. There are several structural 
compatibility conditions which have to be fulfilled to meet that requirement, such as:
\begin{itemize}
 \item $\mathscr{S}(\Mb)$ is a convex set, i.e.\ if $\omega_1,\ldots,\omega_n$ are finitely many elements of 
 $\mathscr{S}(\Mb)$, and $\lambda_1,\ldots,\lambda_n \ge 0$, with $\sum_{n=1}^n \lambda_j = 1$, then also 
 $\overline{\omega} = \sum_{j =1}^n \lambda_j \omega_j$ is contained in $\mathscr{A}(\Mb)$.
 \item If $\omega$ is in $\mathscr{S}(\Mb)$, then every $\Att \in \mathscr{A}(\Mb)$ with $\omega(\Att^*\Att) > 0$
 induces a transformed state $\omega_{\Att}(\Btt) = \omega(\Att^* \Btt \Att)/\omega(\Att^* \Att)$, and it is 
 required that $\omega_{\Att}$ is again in $\mathscr{S}(\Mb)$.
 \item For every arrow $\psi : \Mb \to \widetilde{\Mb}$ in $\mathbb{GH}${\scriptsize$\mathbb{YP}$}, it should 
 hold that $\mathscr{A}(\psi)^* \tilde{\omega} \in \mathscr{S}(\Mb)$ whenever $\tilde{\omega} \in \mathscr{S}(\widetilde{\Mb})$, where $\mathscr{A}(\psi)^*\tilde{\omega} = \tilde{\omega} \circ \mathscr{A}(\psi)$ is the 
 dual map of $\mathscr{A}(\psi)$. 
 \item It is required that $\mathrm{rce}_{\boldsymbol{M}}[h]^*(\mathscr{S}(\Mb)) \subset \mathscr{S}(\Mb)$ where
 $\mathrm{rce}_{\boldsymbol{M}}[h]^*\omega = \omega \circ \mathrm{rce}_{\boldsymbol{M}}$ is the dual of the 
 relative Cauchy evolution. 
 \item Locally, the state space should be a single folium of any of its states. That means, if 
 $\iota_{M,O} :(O,g|_O) \to (M,g)$ is an arrow in $\mathbb{GH}${\scriptsize$\mathbb{YP}$} by identical embedding
 as before, where $\overline{O}$ is a \textit{compact}
 subset of $M$, then 
 $$ \alpha_{M,O}^*\mathscr{S}(M,g) = \mathrm{Fol}(\pi_\omega \circ \alpha_{M,O}) $$
 is to hold for any $\omega \in \mathscr{S}(M,g)$. ($\alpha_{M,O} = \mathscr{A}(\iota_{M,O})$ as defined previously.)

\end{itemize}
Additionally, it is commonly demanded that the state space fulfills  the \textit{split property}, as well as
\textit{local definiteness}, together with the condition of a \textit{stable scaling limit} at every spacetime point.
The latter two conditions imply that the local von Neumann algebras in the GNS representations of any state in the 
state space are of type $\textrm{III}_1$. We will not discuss these topics here any further and refer to 
Fewster (2016) and Fewster and Verch (2015) for further discussion and literature. However, we should like to point out that 
all these properties are fulfilled for the local covariant quantized Klein-Gordon field by defining 
$\mathscr{S}(\Mb)$ as the set of states which lie locally in the folium of any quasifree \textit{Hadamard state}, or equivalently, any quasifree 
state fulfilling the \textit{microlocal spectrum condition}. Again, we refer to Fewster and Verch (2015) for a fuller discussion, and additional references. 
\\[6pt]
In conclusion, while it may not be feasible to single out distinguished states for a local covariant QFT, criteria for 
a state space with structural properties expected for physical states (including a version of local covariance) can be delineated, and these properties are fulfilled for local folia of states for the quantized Klein-Gordon field selected according to the microlocal spectrum condition. This provides a promising guideline for selecting physical state spaces for more general local covariant QFTs, including interacting QFTs, as the microlocal spectrum condition generalizes quite naturally to general QFTs, and is intrinsically local covariant. 
\\ \\
{\small
\textbf{References}
\\[6pt]
Rejzner, K., \textit{Perturbative Algebraic Quantum Field Theory}, Springer-Verlag, 2016
\\[4pt]
Wald, R.M., \textit{Quantum Field Theory in Curved Space Time and Black Hole Thermodynamics}, University of Chicago Press, 1994
\\[4pt]
Brunetti, R., Fredenhagen, K., Verch, R. (2003), ``The generally covariant locality principle -- A new paradigm for local quantum physics'',
Commun.\ Math.\ Phys.\ \textbf{237}, 31–68
\\[4pt]
Fewster, C.J. (2016), ``The split property for quantum field theories in flat and curved spacetimes'',
    Abh.\ Math.\ Sem.\ Univ.\ Hamburg \textbf{86} 2, 153-175
\\[4pt]
Fewster, C.J., Verch, R. (2015), 
``Algebraic quantum field theory in curved spacetimes'', in: R.\ Brunetti, C.\ Dappiaggi, K.\ Fredenhagen, J.\ Yngvason, eds.,
\textit{Advances in Algebraic Quantum Field Theory}, Springer-Verlag
\\[4pt]
Hollands, S. (2008), ``Renormalized quantum Yang-Mills fields in curved spacetime'',
    Rev.\ Math.\ Phys.\ \textbf{20}, 1033-1172
\\[4pt]
Hollands, S., Wald, R.M. (2015), ``Quantum fields in curved spacetime'', Phys.\ Rept.\ \textbf{574}, 1-35
\\[4pt]
Meda, P., Pinamonti, N., Siemssen, D. (2021), ``Existence and uniqueness of solutions of the semiclassical Einstein equation in cosmological models'',
Ann.\ H.\ Poincar\'e \textbf{22}, 3965-4015
\\[4pt]
Sanders, K. (2010),``The locally covariant Dirac field'', Rev.\ Math.\ Phys.\ \textbf{22}, 381-430
\\[4pt]
Sanders, K., Dappiaggi, C., Hack, T.-P. (2014), ``Electromagnetism, local covariance, the Aharonov-Bohm effect and Gauss' law'',  Commun.\ Math.\ Phys.\ \textbf{328}, 625-667
\\[4pt]
Verch, R. (2001), ``A spin-statistics theorem for quantum fields on curved spacetime manifolds in a generally
covariant framework'', Commun.\ Math.\ Phys.\ \textbf{223}, 261-288
\\[4pt]
Wald, R.M. (1979), ``Existence of the $S$-matrix in quantum field theory in curved space-time'',
Ann.\ Phys. (NY) \textbf{118}, 490-510
${}$\\[10pt]
\textbf{Further reading}
\\[6pt]
Buchholz, D., Fredenhagen, K. (2020), ``A $C^*$-algebraic approach to interacting quantum field theory'',
Commun.\ Math.\ Phys.\ \textbf{377}, 947-969
\\[4pt]
Bunk, S., MacManus, J., Schenkel, A. (2025), ``An equivalence theorem for algebraic and functorial QFT'',
    arXiv:2504.15759 [math-ph]
\\[4pt]
Dedushenko, M. (2023),  ``Snowmass white paper: The quest to define Quantum Field Theory'',
 International Journal of Modern Physics \textbf{38}, 2330002 
\\[4pt]
Fewster, C.J., Verch, R. (2012), ``Dynamical locality and covariance: what makes a physical theory the same in all spacetimes?'', Ann.\ H.\ Poincar\'e \textbf{13}, 1613-1674
\\[4pt]
Fewster, C.J., Verch, R. (2013), ``The necessity of the Hadamard condition'', Class.\ Quant.\ Grav.\ \textbf{30},
235027
\\[4pt]
Janssen, D.W. (2022), ``Quantum fields on semi-globally hyperbolic spacetimes'', 
    Commun.\ Math.\ Phys.\ \textbf{391}, 669-705
\\[4pt]
Kay, B.S. (2025), ``Quantum field theory in curved spacetime (2nd edition)'', in: M.\ Bojowald, R.J.\ Szabo, eds., \textit{Elsevier Encyclopedia of Mathematical Physics (Second Edition)} 5, 357-381 (arXiv:2308.14517 [gr-qc])
\\[4pt]
Khavkine, I., Moretti, V. (2015), ``Algebraic QFT in curved spacetime and quasifree Hadamard states: An introduction'',
in: R.\ Brunetti, C.\ Dappiaggi, K.\ Fredenhagen, J.\ Yngvason, eds.,
\textit{Advances in Algebraic Quantum Field Theory}, Springer-Verlag
\\[4pt]
Witten, E. (2022), ``Why Does Quantum Field Theory in Curved Spacetime Make Sense? And What Happens to the Algebra of Observables in the Thermodynamic Limit?'', in:  Mo-Lin Ge, Yang-Hui He (eds.), \textit{Dialogues Between Physics and Mathematics}, Springer-Verlag 
}

\newpage \noindent
\section{Temperature and Entropy-Area Relation of Quantum Fields near Horizons of Dynamical Black Holes}

\subsection{Black Hole Thermodynamics and Hawking effect}

The material in this section is primarily taken from the article Kurpicz, Pinamonti and Verch (2021) referenced at the end of the section.
See also the remarks about related work towards the end of this section.
\\[6pt]
The laws of {\bf black hole thermodynamics} by Bardeen, Carter and Hawking (1973) for \underline{stationary} black hole spacetimes: 
\begin{itemize}
 \item  {\bf zero}th {\bf law}: \ \
 The {\bf surface gravity} $\kappa$ is constant on the horizon
 \item {\bf first law}: 
 $$ d\mathcal{M} = \frac{\kappa}{8 \pi} dA + \Omega dJ \ \ \left( \, + Q d\Phi \right) $$
 \item{\bf second law}: 
 $$ \frac{dA}{dt} \ge 0 $$
\end{itemize}
$\mathcal{M}$ = black hole mass(-energy), $A$ = black hole horizon surface area,
$J$ = black hole angular momentum, $\Omega$ = black hole angular velocity.
\\[2pt]
Here,
$t$ is a suitable time-coordinate with $\partial/\partial t$ future-pointing, thus the third law says that the horizon surface area of a black hole can never shrink. Interestingly, this apparently means 
that black holes can only grow and not shrink; in particular, according to the third law, black holes cannot ``split'' into smaller fragments (at least this seems to be the general interpretation). 
\\[6pt]
These considerations are based entirely within classical spacetime geometry. Physical units are chosen so that  $G = 1$, $c = 1$.
\\[6pt]
Bardeen, Carter and Hawking (1973) have suggested an analogy with thermodynamical quantities --
\begin{align*}
{\bf stationary\ black\ hole}     \quad \ \ & \quad \ \ \ \ \quad {\bf system\ in\ thermal\ equilibrium} \\[8pt]
  \text{mass-energy} \ \ \mathcal{M} \ \ & \longleftrightarrow \ \ \mathcal{E} \ \ \text{internal energy} \\[6pt]
  \text{surface gravity} \ \ \kappa \ \ & \longleftrightarrow \ \ \frac{1}{T} \ \ \text{inverse temperature} \\[6pt]
  \text{surface area} \ \ A \ \ & \longleftrightarrow \ \ S \ \ \text{entropy} \\[6pt]
  \text{angular velocity} \ \ \Omega \ \ & \longleftrightarrow \ \ p \ \ \text{pressure} \\[6pt]
  \text{angular momentum} \ \ J \ \ & \longleftrightarrow \ \ V \ \ \text{volume}
\end{align*}
The (microscopic) origin of ``black hole temperature'' can be explained within {\bf QFT in curved spacetime},
as has been done in the seminar work
Hawking (1975) on the thus named \textit{Hawking effect}. This effect is based on a quantum field scattering scenario on a spacetime with classical matter background collapsing to a Schwarzschild black hole.
\begin{itemize}
 \item 
\underline{initially} (at very early time), matter is thinly spread, metric approximately flat, QFT state approximately vacuum
\item
\underline{finally} (at very late time), matter has collapsed to Schwarzschild black hole, QFT state is a KMS state with respect to Schwarzschild time outside of the event horizon
at inverse temperature $\beta = 2\pi/\kappa$,\\ $\kappa =$ surface gravity of horizon ($\sim 1/R_S$)
\end{itemize}
Several variants and improvements of these arguments have since appeared. See the suggestions for Further Reading at the end of the section. Of some interest is the work of Sewell (1982) who has pointed out that there is a setting similar 
to that of Bisognano-Wichmann for QFT on the outside spacetime region of a static black hole, by which the 
``natural'' state of a quantum field on the outside region of a static black hole is a KMS state with respect 
to Schwarzschild time, at an inverse temperature that matches the Hawking temperature. See also Kay and Wald (1991) for 
related results.
\\[6pt] All these considerations use the {\bf global} spacetime structure. On the other hand,
a {\bf local} concept of Hawking temperature is desirable, e.g.\ to describe ``black hole evaporation'' in semi-classical gravity.
\\[6pt]
Parikh-Wilczeck (2000): ``Tunneling'' interpretation using a quantum mechanics inspired approach ( + very many follow-up papers)
\\[6pt]
Moretti-Pinamonti (2012): ``Tunneling'' interpretation in a QFT scaling limit for a stationary black hole horizon
\\[10pt]
What is the interpretation of ``black hole entropy'' and the  entropy-area relation, and to which (microscopic) degrees of freedom does it refer? In 
Bekenstein (1973), it was suggested that black hole entropy corresponds to the information lost about states falling into the black hole, and an  illustration was given using light beams crossing into the black hole horizon. Several
other ideas on black hole entropy have been discussed. Worth mentioning is the argument of
Wald (1993) stating that black hole entropy is Noether charge. Other concepts of black hole entropy have 
appeared in relation to loop quantum gravity and to string theory, and to holography. 
We won't discuss any of these approaches and instead will try to understand black hole entropy in the conceptual
framework of quantum field theory in curved spacetime. Recently, there has been some new development in this direction,
to some degree based on the insight that relative entropy is a suitable quantity to investigate in that context, allowing to derive new results (see, e.g.\ suggestions for Further Reading). The approach presented here also follows similar lines.

\subsection{Spacetimes with outer trapping horizon: Dynamical black holes}

We follow Hayward (1998) and Ashtekar et al.\ (2004) for the definition of a spherically symmetric \textbf{spacetime with outer trapping horizon (OTH)}.
\begin{itemize}
 \item manifold: $M = \mathcal{L} \times S^2$ \quad \ \ \ $(\mathcal{L} \subset \mathbb{R}^2)$
 \\[6pt]
 \item metric $g_{ab}$, \hfill [$(-\, +\, +\, +)$]
 \begin{align*}
 ds^2 & =   -e^{2\Psi(v,r)}C(v,r)dv^2  +   2e^{\Psi(v,r)} dv dr    + r^2 d\Omega^2 \\
  & \ d\Omega^2 = \sin^2(\vartheta)d\varphi^2 + d\vartheta^2\ \ \text{(spherically symmetric)}
\end{align*}
\end{itemize}
$\longrightarrow$ generalization of spherically symmetric central mass distribution varying in time,
$$ \mathcal{M} = \frac{r}{2} (1 - C) \quad \ \ \text{Hawking mass} $$
if $\Psi = 0$ $\longrightarrow$ Vaidya metric --- generalization of Schwarzschild metric with time-dependent
central mass (see Griffiths and Podolsky (2010) for discussion).
\\[10pt]
{\bf Definition}
$$ \mathcal{H} = \{ (v,r,\vartheta,\varphi) : C(v,r) = 0 \} \quad \ \ \mathbf{outer\ trapping\ horizon} $$
\begin{itemize}
\item $\mathcal{M}$ decreasing in time $\longrightarrow$ ``black hole evaporation''
$\longrightarrow$ $\mathcal{H}$ is \textsl{timelike}
\item $\mathcal{M}$ increasing in time $\longrightarrow$ ``black hole mass accretion'' 
$\longrightarrow$ $\mathcal{H}$ is \textsl{spacelike}
\item $\mathcal{M}$ constant in time $\longrightarrow$ $\mathcal{H}$ is \textsl{lightlike}
\end{itemize}
${}$\\
$\mathcal{H}$ separates the part of spacetime where radially outward directed lightlike geodesics have positive expansion (``outside'') from the part where
radially outward directed lightlight geodesics have negative expansion (``inside'').
${}$ \\
\begin{center}
 \includegraphics[width=9.5cm]{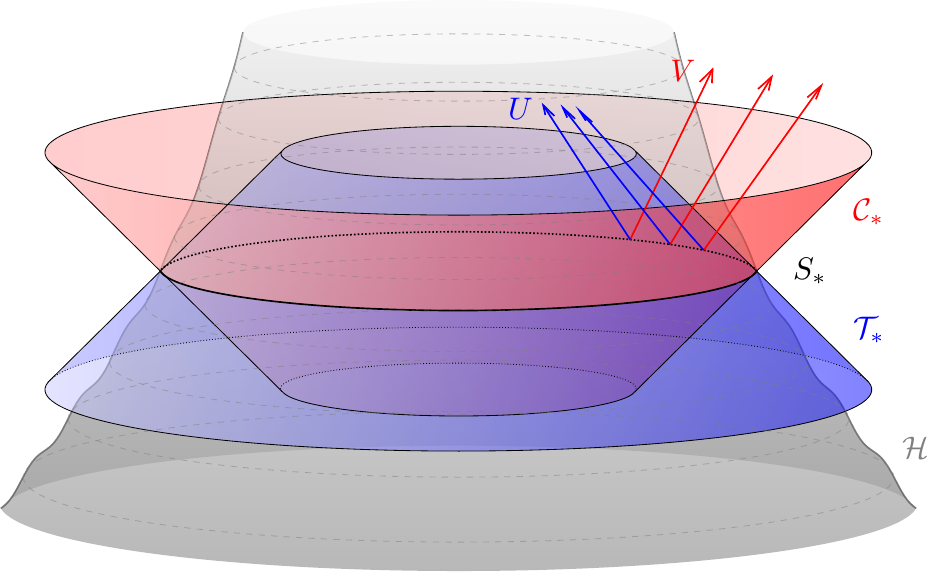}
\end{center}
{\small
\textbf{Figure 9.} Illustration of the lightlike hypersurfaces $\mathcal{C}_*$ (outgoing lightlike congruence, shaded red) and $\mathcal{T}_*$ (ingoing lightlike congruence, shaded blue) emanating from
a horizon cross-section $S_*$, i.e.\ an orbit of a point on $\mathcal{H}$ (shaded grey) under the action of the group of rotations. The lightlike hypersurfaces are here drawn as they would appear in an ambient Minkowski spacetime, with $45$ degrees inclination against the vertical coordinate axis. Therefore, the lightlike
geodesic generators of $\mathcal{C}_*$ appear to have positive expansion, i.e.\ to drift apart towards the future. According to the definition of the OTH, the lightlike geodesic generators of $\mathcal{C}_*$ have zero expansion, which 
would be achieved by drawing $\mathcal{C}_*$ as the surface of an upright cylinder with cross-section $S_*$. However,
this would not invoke the impression of a lightlike hypersurface --- we caution readers about the limitations of the 
illustration. $U$ and $V$ are affine parameters for the lightlike geodesic generators of $\mathcal{T}_*$ and $\mathcal{C}_*$ in adapted coordinates, described below. Figure generated by F.\ Kurpicz.}
\\ \\
Spacetimes with OTH might not admit any timelike Killing vector field --- but a generalization:
The \textbf{Kodama vector field} (Kodama (1980))
$$K^a = {\rm e}^{-\Psi} \frac{\partial}{\partial v}^a $$
The Kodama vector field has the properties $\nabla_a K^a = 0$ and 
$\nabla_a (W^{ab}K_b) = 0$ if $W^{ab}$ is any rotation-invariant and symmetric degree 2 co-tensorfield.
\\[6pt]
The \textbf{surface gravity} $\kappa$ on $\mathcal{H}$ is defined by
$$ \kappa = \frac{1}{2}\frac{\partial C}{\partial r} \quad \Longleftrightarrow \quad  \frac{1}{2}K^a (\nabla_a K_b-\nabla_b K_a) = \kappa K_b  \quad \text{on} \ \ \mathcal{H}
 $$
\textbf{Adapted coordinates} \ \  $(U,V,\vartheta,\varphi)$ \ \ near $\mathcal{H}$ \\[2pt]
relative to a point $(v_*,r_*,\vartheta_*,\varphi_*)$ on $\mathcal{H}$, resp.\ its rotation-orbit
$S_* = {(v_*,r_*)} \times S^2$\,:
\\[4pt]
$V$ is affine parameter for the {\bf outgoing lightlike geodesic congruence} $\mathcal{C}_*$,
\\[4pt]
$U$ is affine parameter for the {\bf ingoing lightlike geodesic congruence} $\mathcal{T}_*$,
\\[4pt]
each {\bf emanating from $S_*$} (cf.\ Figures 9 and 10)
\\[10pt]
${}$ \quad 
(1) $U = 0$ and $V = 0$ exactly for the points on $S_*$, 
\\[2pt]
${}$ \quad 
(2) 
$U = 0$ exactly for the points on $\mathcal{C}_*$, 
\\[2pt]
${}$ \quad 
(3) $V = 0$ exactly for the points on $\mathcal{T}_*$, 
\\[2pt]
${}$ \quad 
(4)
$ds^2 = -2A(U,V)dU\,dV + r^2(U,V) d\Omega^2$ with
 $A = 1$ on $\mathcal{C}_* \cup \mathcal{T}_*$, 
\\[2pt]
${}$ \quad 
(5) $dU_a K^a (U,V = 0,\vartheta,\varphi) = -\kappa_* U + O(U^2)$ \\[2pt]
${}$ \quad \quad \quad on $\mathcal{T}_*$ near $U = 0$, with $\kappa_* = \left. \kappa\right|_{S_*}$.
${}$ \\ ${}$
\begin{center}
\includegraphics[width=12cm]{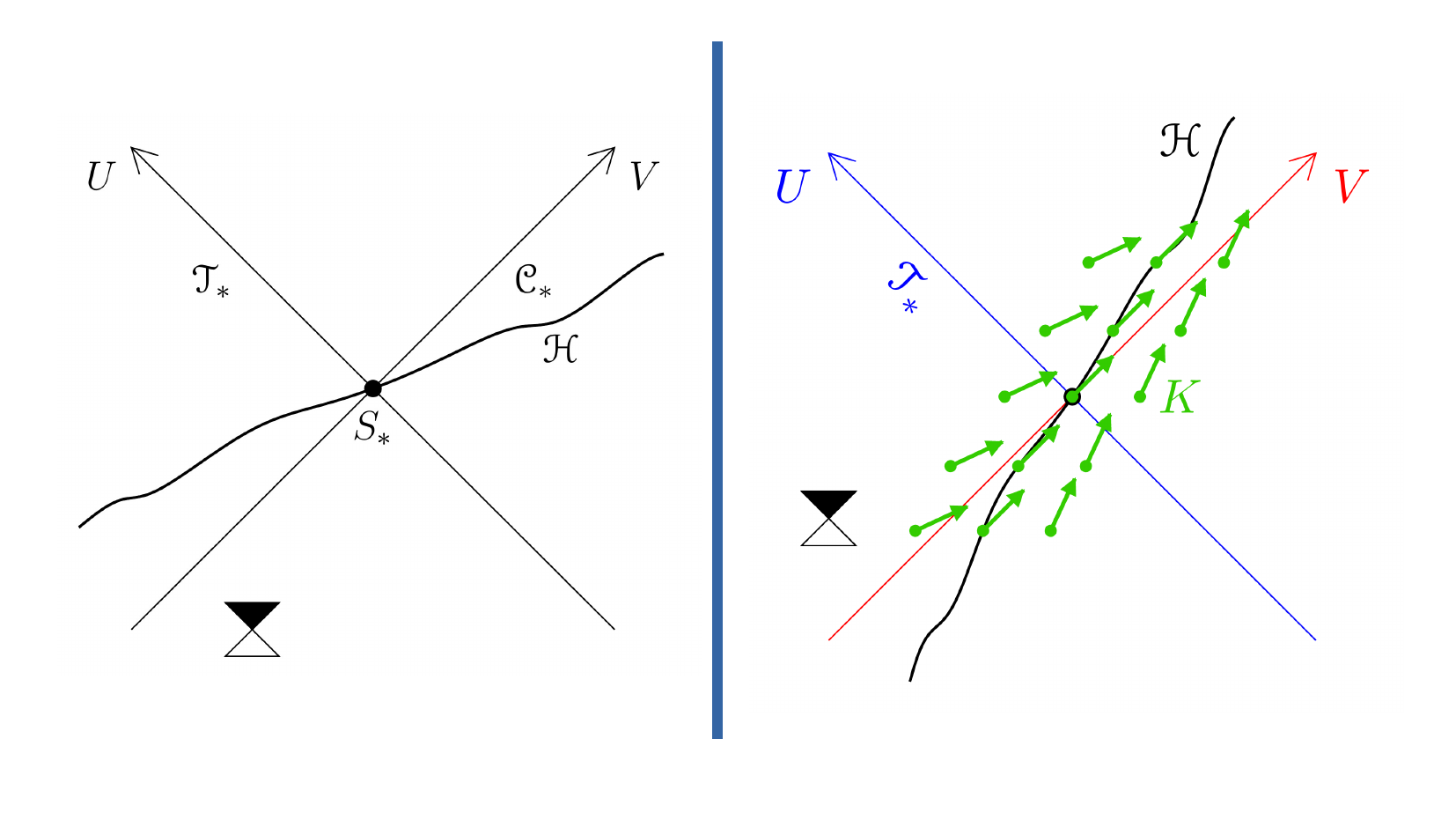}
\end{center}
{\small
\textbf{Figure 10.} Sketch of the outer trapping horizon $\mathcal{H}$ with the cross section $S_*$, represented as a point. The lightlike geodesics generating $\mathcal{T}_*$ and $\mathcal{C}_*$ are indicated as lines
at 45 degrees inclination. The adapted coordinates $U$ and $V$ are indicated. On the left hand side, the 
outer trapping horizon is spacelike. On the right hand side, the outer trapping horizon is timelike. The green
arrows indicate the Kodama vector field. At $S_*$, the Kodama vector field has no component along the $U$-coordinates, corresponding to property (5). Figure generated by F.\ Kurpicz.
}
\\[6pt]
It is also assumed that the spacetime $(M,g_{ab})$ is globally hyperbolic (if needed, restrict to suitable neighbourhood of $\mathcal{H}$ or of parts of it)

\subsection{Quantized linear Klein-Gordon field} 
\begin{itemize}
 \item $\mathcal{A} = \mathcal{A}(M,g_{ab},{\rm M})$ is the unital $*$-algebra generated by a unit $\mathbf{1}$ together with elements 
 $$ \Phi(F)\,, \quad F \in C_0^\infty(M,\mathbb{R}) \quad \quad \text{abstract field operators; relations:} $$
 \item $\Phi((\nabla^a \nabla_a + {\rm M})F) = 0$;  \quad ${\rm M} \in C^\infty(M,\mathbb{R})$ \ \ potential function 
 \item $[\Phi(F_1),\Phi(F_2)] = i (G_+(F_1,F_2) - G_-(F_1,F_2)){\bf 1}$ \,, \\[4pt]
 $G_\pm$ are the {\bf unique advanced/retarded Green-distributions} of\ \  $\nabla^a \nabla_a + {\rm M}$ 
\end{itemize}

States are linear, positive, normalized functionals  $ \omega :\mathcal{A} \to \mathbb{C}$ determined by their $n$-point distributions
$$ w^{(n)}(F_1 \otimes \cdots \otimes F_n) = \omega(\Phi(F_1) \cdots \Phi(F_2)) \,, \quad \ \ F_j \in C_0^\infty(M,\mathbb{R})$$
Quasifree states are entirely determined by their 2-point distribution $w^{(2)}$. A 2-point distribution is said to 
be
of \textbf{Hadamard form} if 
\begin{align*}
 w^{(2)}(F,F') = \lim_{\varepsilon \to 0+} \int w_\epsilon(x,x') F(x)F'(x')
\,d{\rm vol}_g(x)\,d{\rm vol}_g(x') 
\end{align*}
with integral kernels for $\varepsilon > 0$ of the form
\begin{align*}
  w_\varepsilon(x,x') =   \frac{\Delta^{1/2}(x,x')}{8\pi^2\, \sigma_\varepsilon(x,x')} 
  +\  \ln(\sigma_\varepsilon(x,x'))Y(x,x')  + Z(x,x')
\end{align*}
where $\sigma_\varepsilon(x,x')$ is a ``regularized'' squared geodesic distance between
the spacetime points $x$ and $x'$ which doesn't become 0 for $\varepsilon > 0$, and approaches the squared geodesic distance as 
$\varepsilon \to 0$. The quantities $\Delta^{1/2}(x,x')$ and $Y(x,x')$ are smooth and locally determined by 
the spacetime metric. This implies that the singular part of a Hadamard-type 2-point distribution is entirely 
determined by the spacetime metric. There is a smooth term $Z(x,x')$ which carries the state dependence. 
For further details not spelled out here, we refer to Kay and Wald (1991) and Khavkine and Moretti (2015). 
We also mention that the Hadamard form of the 2-point distribution is equivalent to requiring that the wavefront set 
of the 2-point distribution is contrained in a particular way; that latter condition is now commonly termed
``microlocal spectrum condition''. See Radzikowski (1996) and Khavkine and Moretti (2015) as well as references
cited there for considerably more on these matters.

\subsection{Scaling limit of a Hadamard state at a chosen $S_*$}

We aim at investigating the behaviour of the quantized Klein-Gordon field near the OTH $\mathcal{H}$. We will do this by
carrying out a short-distance scaling limit of the quantized Klein-Gordon field at a given $S_* \subset \mathcal{H}$, and 
analyzing the properties of the scaling limit. We will see that the scaling limit can be interpreted as describing
a quasifree vacuum state of a quantum field theory on the lightlike hypersurface $\mathcal{T}_*$ of ingoing lightrays emanating 
from $S_*$, for which a Borchers-triple structure emerges naturally. The latter will be instrumental in deriving, in the scaling limit,
an entropy-area relation. The procedure is as follows:
\\[8pt]
1. Choose $S_*$ \\[4pt]
2. Set up adapted coordinates $(U,V,\vartheta,\varphi)$ w.r.t. $S_*$\\[4pt]
3. Define transformations on test-functions: \hfill ($0 < \mu,\lambda < 1$)
\begin{align*}
 ({\sf u}_\lambda F)(U,V,\vartheta,\varphi) & = \frac{1}{\lambda}F(U/\lambda,V,\vartheta,\varphi) \quad  (\text{scaling towards} \ U = 0) \\[10pt]
 ({\sf v}_\mu F)(U,V,\vartheta,\varphi) & = \underset{\delta-\text{sequence}}{\underbrace{\frac{1}{\mu} \zeta (V/\mu)}} F(U,V,\vartheta,\varphi) 
 \quad (\text{restriction onto} \ V = 0)
\end{align*}
Let $w^{(2)}$ be a 2-point function of Hadamard form on the spacetime with OTH. \\[6pt]
There is an open neighbourhood $\mathcal{O}$ of $S_*$ so that
for all $f,f' \in C_0^\infty(\mathcal{O},\mathbb{R})$ --- 
\begin{align*}
\lim_{\mu \to 0} & \lim_{\lambda \to 0}\,w^{(2)}({\sf v}_\mu{\sf u}_\lambda({2}\partial_U f),{\sf v}_\mu{\sf u}_{\lambda}({2}\partial_{U'}f'))
 = \Lambda(f,f')
\\[6pt]
= \ \lim_{\varepsilon\to 0+} &  -\frac{r_*^2}{{\pi}} \int  \frac{ f(U,0,\vartheta,\varphi) f'(U',0,\vartheta,\varphi)}{(U-U'+i\varepsilon)^2}
dU\,dU'\,d\Omega^2
  \end{align*}
Denoting by $\mathcal{K}_t$ $(t \in\mathbb{R})$ the \textbf{flow} of the Kodama vector field, it holds that 
$$ \lim_{\mu \to 0} \,\lim_{\lambda \to 0}\,w^{(2)}(\mathcal{K}_t^*{\sf v}_\mu{\sf u}_\lambda({2}\partial_U f),\mathcal{K}_{t'}^*{\sf v}_\mu{\sf u}_{\lambda}({2}\partial_{U'}f'))
 = \Lambda(\tau_t f,\tau_{t'} f')\,, $$
$$ \text{where} \ \ \tau_tf(U,V,\vartheta,\varphi) = f({\rm e}^{\kappa_*t}U,V,\vartheta,\varphi) $$
Through the scaling limit and restriction, independently of the Hadamard 2-point function chosen, a {\bf scaling limit theory} is defined relative to any 
$S_* \subset \mathcal{H}$:\\[4pt]
A conformal QFT on $\mathbb{R} \times (S^2 \simeq S_*) \supset \mathcal{T}_* \cap \mathcal{O}$,
based on the {\bf vacuum 2-point function} $\Lambda$.
\\[6pt]
From now on: Identify \ \
$f(U,0,\vartheta,\varphi) \leftrightarrow f(U,\vartheta,\varphi)$, $f \in C_0^\infty(\mathbb{R} \times S^2,\mathbb{R})$
\\[6pt]
\textbf{Symplectic form:} \quad  $\varsigma(f,f') = 2 {\rm Im}\,\Lambda(f,f')$
\\[6pt]
\textbf{Weyl-algebra $\mathcal{W}_{S_*}$:} \quad generators $W(f)$, $f \in C_0^\infty(\mathbb{R} \times S^2,\mathbb{R})$,
$$
W(0) = {\bf 1}\,, \quad W(f)^* = W(-f)\,, \quad W(f)W(f') = {\rm e}^{{ -\frac{i}{2}}  \varsigma(f,f')} W(f + f')
$$
\textbf{Local algebras:} \quad  $\mathcal{W}_{S_*}(G)$ generated by $W(f)$, ${\rm supp}(f) \subset G \subset \mathbb{R} \times S^2$ \\[6pt]
\textbf{Vacuum state $\omega_\Lambda$:} \quad quasifree state induced by 
$$ \omega_\Lambda(W(f)) = {\rm e}^{-\Lambda(f,f)/2} $$
\textbf{Local vN algebras in GNS representation} $(\mathcal{H}_\Lambda,\pi_\Lambda,\Omega_\Lambda)$ of $\omega_\Lambda$:
$$ \mathsf{N}(G) =  \pi_\Lambda(\mathcal{W}_{S_*}(G))'' \ \subset \ \mathsf{B}(\mathcal{H}_\Lambda) $$
\textbf{Unitary group representation}\ \ ${\rm U}(t,a,R)$ of dilations, translations and rotations:\\[2pt]
$\Lambda$ is invariant under the action of the group formed by dilations, translations and rotations: For all $f, h \in C_0^\infty(\mathbb{R} \times S^2,\mathbb{R})$,
it holds that 
$$ \Lambda(f_{(t,a,R)},h_{(t,a,R)}) = \Lambda(f,h) \quad \text{where} \quad 
f_{(t,a,R)}(U,\vartheta,\varphi) = f({\rm e}^{\kappa_*t}U - a,R^{-1}(\vartheta,\varphi)) $$
for $t,a \in \mathbb{R}$, $R \in SO(3)$. Consequently, in the GNS representation of $\omega_\Lambda$ there is a 
continuous unitary representation $\mathrm{U}(t,a,R)$ of the group of dilations, translations and rotations, such that
$$ {\rm U}(t,a,R)\pi_\Lambda(W(f)){\rm  U}(t,a,R)^* = \pi_\Lambda(W(f_{(t,a,R)})) \,, \quad  {\rm U}(t,a,R)\Omega_\Lambda = \Omega_\Lambda$$
Here $\kappa_* = \left. \kappa \right|_{S_*}$
\\[10pt]
Setting $\mathsf{N}_R = \mathsf{N}((-\infty,0) \times S^2)$, $\mathsf{N}_L = \mathsf{N}((0,\infty) \times S^2)$,
one obtains a \textbf{geometric action of modular objects} as in the Bisgnano-Wichmann/Borchers theorem:
\begin{itemize}
 \item $\Omega_\Lambda$ is a cyclic and separating vector for both $\mathsf{N}_{L/R}$
 \item translation group ${\rm U}(a)$ has positive generator and ${\rm U}(a) \mathsf{N}_R{\rm  U}(-a) \subset \mathsf{N}_R$ $(a \le 0)$
 \item dilation group ${\rm U}(t)$ is given by the {\it Tomita-Takesaki modular group} $\Delta^{it}$ of $\mathsf{N}_R$ and $\Omega_\Lambda$:
 $$ \Delta^{it/\beta}= {\rm U}(t)\,, \ \ \ \beta = 2\pi/\kappa_* \quad \ \ (t \in \mathbb{R}) $$
equivalently, $\omega_\Lambda$ is a {\bf KMS-state} at inverse temperature $\beta = 2\pi/\kappa_*$ with respect to the 
dilations along $\mathcal{T}_*$ arising from the Kodama flow in the scaling limit.
\\[6pt]
This is {\bf a local analogue of the Hawking effect} in the scaling limit at $S_*$. 
\end{itemize}

\subsection{Interpretation as ``tunneling probability''}

Let $f_{L,n} \in C_0^\infty((0,\infty)\times S^2,\mathbb{R})$, $f_{R,n} \in C_0^\infty((-\infty,0) \times S^2,\mathbb{R})$ $(n \in \mathbb{N})$
so that their {\bf Fourier-transforms with respect to the coordinate} $t$ (Kodama-flow parameter in the scaling limit) approach 
$\delta_{E_0}$ (as $n \to \infty$) for some $E_0 > 0$; then normalize the functions ---
$$ \psi_{L/R,n} = \frac{f_{L/R,n}}{\Lambda(f_{L/R,n},f_{L/R,n})^{1/2}} $$
Then $\psi_{L/R,n}$ represent unit-norm one-particle vectors in $\mathcal{H}_\Lambda$, and they have the property
\begin{align*}
|\langle \psi_{L,n},\psi_{R,n}\rangle |^2 = |\Lambda(\psi_{L,n},\psi_{R,n})|^2 \ \ 
\underset{n\to \infty}{\longrightarrow} \ \ \frac{1}{4}\left(\frac{1 - {\rm e}^{-\beta E_0}}{\sinh(\beta E_0/2)}\right)^2 \approx {\rm e}^{-\beta E_0}
\end{align*}
i.e.\ the ``transition probability'' between scaling limit modes inside (L) and outside (R) of the OTH shows a thermal energy distribution for large 
$E_0$ at inverse temperature $\beta = 2\pi/\kappa_*$

\subsection{Relative entropy of coherent states in the scaling limit theory and entropy-area relation}

{\bf Coherent states} of the scaling limit theory: If $f \in C_0^\infty((-\infty,0) \times S^2, \mathbb{R})$
$$ \omega_f^{\rm coh}(A) = \omega_\Lambda(W(f) A W(f)^*) \quad \ \ (A \in \mathsf{N}_R) $$
The {\bf relative entropy} (Araki (1976), Uhlmann (1977)) between the coherent states and the scaling limit state $\omega_\Lambda$ can be calculated as (cf.\ Hollands (2019), Longo (2019))
\begin{align*}
 {\rm S}(\omega_\Lambda | \omega^{\rm coh}_f) & = i \left. \frac{d}{dt}\right|_{t=0}
 \langle \pi_\Lambda(W(f))\Omega_\Lambda,\Delta^{it} \pi_\Lambda(W(f))\Omega_\Lambda \rangle
\\[6pt]
= & {-2 }\pi r_*^2 \int_{(-\infty,0)\times S^2} U (\partial_U f)^2(U,\vartheta,\varphi) \,dU\,d\Omega^2(\vartheta,\varphi)
\end{align*}
Here $r_* = \left. r \right|_{S_*}$, so {\bf the relative entropy of coherent states scales like the geometric area $4\pi r_*^2$ of $S_*$}.
\\[6pt]
Reason for entropy-area proportionality:
\\[4pt]
The coherent states are {\bf correlation-free across disjoint supports on $S^2$}\\[4pt]
$\Longrightarrow$ \quad relative entropy is additive in the sense:
\\[4pt]
If the $S^2$-supports of $f_1$ and $f_2$ are disjoint, 
\begin{align*}
 {\rm S}(\omega_\Lambda | \omega^{\rm coh}_{f_1 + f_2}) & = {-2 }\pi r_*^2 \int_{(-\infty,0)\times S^2} U (\partial_U (f_1 +f_2))^2(U,\vartheta,\varphi) \,dU\,d\Omega^2(\vartheta,\varphi)
 \\
 & = \sum_{j = 1,2} {-2 }\pi r_*^2 \int_{(-\infty,0)\times S^2} U (\partial_U f_j)^2(U,\vartheta,\varphi) \,dU\,d\Omega^2(\vartheta,\varphi) \\
& =  {\rm S}(\omega_\Lambda | \omega^{\rm coh}_{f_1}) + {\rm S}(\omega_\Lambda | \omega^{\rm coh}_{f_2}) 
\end{align*}

In particular, consider
\begin{align*}
 (h \odot \chi_\Sigma)(U,\vartheta,\varphi) = h(U) \cdot \chi_\Sigma(\vartheta,\varphi)  \quad (U \in (-\infty,0)\,, \ (\vartheta,\varphi) \in S^2)
\end{align*}
with $h \in C_0^\infty((-\infty,0),\mathbb{R})$ and 
$\chi_{\Sigma}$ characteristic function of $\Sigma \subset S^2_{r_*} \equiv S_*$, then the following obtains:
\begin{align*}
{\rm S}(\omega_\Lambda| \omega^{\rm coh}_{h \odot \chi_\Sigma}) & = -{2} \pi  \int_{-\infty}^0 U (\partial_U h)^2(U) \, dU \cdot 
 r_*^2\int_{\Sigma \subset S^2} d\Omega^2(\vartheta,\varphi) \nonumber \\
 & = - {2} \pi \int_{-\infty}^0 U (\partial_U h)^2(U) \, dU \cdot {\rm Area}(\Sigma \subset S_*) 
\end{align*}
This shows that the relative entropy of the quantum field coherent states is additive in the horizon 
surface area. 
\\[6pt]
We mention that the connection between a geometric action of Tomita-Takesaki modular objects and entropy in the context of black holes, as well as the connection to certain quantum energy conditions, has developed into an important branch in the mathematical research on quantum field theory over the past several years. An appropriate discussion of that topic would deserve at least a chapter in its own right, but we shall not embark on these matters, and refer to D'Angelo (2021) and Hollands and Longo (2025) and references given there for more about this current line of research. 
${}$\\[10pt]
{\small
\textbf{References}
\\[6pt]
Griffiths, J.B., Podolsky, J., \textit{Exact Spacetimes in Einstein's General Relativity}, Cambridge University Press, 2010
\\[4pt]
Araki, H. (1976), ``Relative entropy of states of von Neumann algebras'', Publ.\ Res.\ Inst.\
Math.\ Sci.\ \textbf{11}, 809–833.
\\[4pt]
Ashtekar, A., Krishnan, B. (2004), ``Isolated and dynamical horizons and their applications'', Living Rev.\ 
Relativ.\ \textbf{7}, 10 
\\[4pt]
Bardeen, J., Carter, B., Hawking, S.W. (1973), ``The four laws of black hole mechanics'', Commun.\ Math.\ Phys.\  \textbf{31}, 161–170 
\\[4pt]
Bekenstein, J. (1973), ``Black holes and entropy'', Phys.\ Rev.\ \textbf{D 7}, 2333 
\\[4pt]
Hawking, S.W. (1975), ``Particle creation by black holes'', Commun.\ Math.\ Phys.\ \textbf{43}, 199 
\\[4pt]
Hayward, S.A. (1998), ``Unified first law of black-hole dynamics and relativistic thermodynamics'', Class.\ Quant.\
Grav.\ \textbf{15}, 3147–3162 
\\[4pt]
Hollands, S. (2020), ``Relative entropy for coherent states in chiral CFT'',
Lett.\ Math\ .Phys.\ \textbf{110}, 713-733
\\[4pt]
Kay, B.S., Wald, R.M. (1991), ``Theorems on the uniqueness and thermal properties of stationary, nonsingular,
quasifree states on spacetimes with bifurcate Killing horizons'', Phys.\ Rept.\ \textbf{207}, 49–136 
\\[4pt]
Kodama, H. (1980), ``Conserved energy flux for the spherically symmetric system and the backreaction problem
in the black hole evaporation'', Prog.\ Theor.\ Phys.\ \textbf{63}, 1217 
\\[4pt]
Kurpicz, F., Pinamonti, N., Verch, R. (2021),
``Temperature and entropy–area relation of quantum matter near spherically symmetric outer trapping horizons'',
Lett.\ Math.\ Phys.\
\textbf{111}, 110
\\[4pt]
Longo, R. (2019), ``Entropy of coherent excitations'', Lett.\ Math.\ Phys.\ \textbf{109}, 2587-2600
\\[4pt]
Moretti, V., Pinamonti, N. (2012), ``State independence for tunnelling processes through black hole horizons and Hawking radiation'',
Commun.\ Math.\ Phys.\ \textbf{309}, 295–311
\\[4pt]
Parikh, M.K., Wilczek, F. (2000), ``Hawking radiation as tunneling'', Phys.\ Rev.\ Lett.\ \textbf{85}, 5042-5045
\\[4pt]
Radzikowski, M.J. (1996), ``Micro-local approach to the Hadamard condition in quantum field theory on curved space-time'',     Commun.\ Math.\ Phys.\ \textbf{179}, 529-553
\\[4pt]
Sewell, G.L. (1982), ``Quantum fields on manifolds: PCT and gravitationally induced thermal states'', Ann.\ Phys.\
(N.Y.) \textbf{141}, 201 
\\[4pt]
Uhlmann, A. (1977), ``Relative entropy and the Wigner-Yanase-Dyson-Lieb concavity in an
interpolation theory'', Commun.\ Math.\ Phys.\ \textbf{54}, 21–32
\\[4pt]
Wald, R.M. (1993), ``Black hole entropy is Noether charge'', Phys.\ Rev.\ \textbf{D 48}, R3427
\\[10pt]
\textbf{Further Reading}
\\[6pt]
Dappiaggi, C., Moretti, V., Pinamonti, N., \textit{Hadamard States from Lightlike Hypersurfaces}, Springer-Briefs in Mathematical Physics, vol.\ 25, Springer-Verlag, 2017
\\[4pt]
Hollands, S., Sanders, K., \textit{Entanglement Measures and Their Properties in Quantum Field Theory}, Springer-Briefs in Mathematical Physics, vol.\ 34, Springer-Verlag, 2018 
\\[4pt]
Wald, R.M., \textit{General Relativity}, University of Chicago Press, 1984
\\[4pt]
Ciolli, F, Longo, R., Ranallo, A., Ruzzi, G. (2022), ``Relative entropy and curved spacetimes'',
    J.\ Geom.\ Phys.\ \textbf{172}, 104416
\\[4pt]
D'Angelo, E. (2021), ``Entropy for spherically symmetric, dynamical black holes from the
relative entropy between coherent states of a scalar quantum field'', Class.\ Quant.\
Grav.\ \textbf{38} (2021), 175001
\\[4pt]
Fredenhagen, K., Haag, R. (1990), ``On the derivation of Hawking radiation associated with the formation of
a black hole'', Commun.\ Math.\ Phys.\ \textbf{127}, 273–284 
\\[4pt]
Hollands, S., Ishibashi, A. (2019), ``News versus information'', Class.\ Quant.\ Grav.\ \textbf{36}, 195001
\\[4pt]
Hollands, S., Longo, R. (2025), ``A new proof of the QNEC'', arXiv:2503.04651[hep-th]
\\[4pt]
Janssen, D.W., Verch, R. (2023),    ``Hadamard states on spherically symmetric characteristic surfaces, the semi-classical Einstein equations and the Hawking effect'',    Class.\ Quant.\ Grav.\ \textbf{40}, 045002
\\[4pt]
Wald, R.M. (2001), ``The thermodynamics of black holes'', Living Rev.\ Relativ.\ \textbf{4}, 6
}
\end{document}